%% file: 0.tex
\documentclass[amsthm]{elsart}

\usepackage{yjsco}
\usepackage{natbib}
\usepackage{amsmath}
\usepackage{mathrsfs}
\usepackage{amssymb}
\usepackage{latexsym} 
\usepackage{verbatim}
\usepackage{enumerate}
\usepackage{xspace}
\usepackage{url}

\usepackage{hyperref}
\hypersetup{
    colorlinks,%
    citecolor=blue,%
    linkcolor=blue,%
    urlcolor=blue,
}

\newcommand \rrule[2]{#1\rightarrow #2}
\newcommand \free[1]{\mathcal{V}ar(#1)}
\newcommand \match[2]{#1 \ll #2}
\newcommand \matchth[3]{#1 \ll_{#3} #2}
\newcommand \ST {\textsf{symbtrans}\xspace}

\newcommand \OEps {O(\varepsilon)}
\newcommand \OEpsi[1]{O(#1,\varepsilon)}

\newcommand \ucomment[1]{}

\newcommand \some {\texttt{Some}}
\newcommand \STTop {\texttt{Outermost}}
\newcommand \STBottom {\texttt{Innermost}}

\begin{document}
\begin{frontmatter}

\title{A Symbolic Transformation Language\\
 and its Application to a Multiscale Method}

\thanks{This work has been partially funded by the Labex ACTION
(ANR-11-LABX-01-01), the INTERREG IV project OSCAR and the PPF MIDi.}

\author[in]{Walid Belkhir},
\ead{walid.belkhir@inria.fr}
\author[in,di]{Alain Giorgetti},
\ead{alain.giorgetti@femto-st.fr}
\author[tf]{Michel Lenczner}
\ead{michel.lenczner@utbm.fr}

\address[in]{INRIA Nancy - Grand Est, CASSIS project, 54600 Villers-l\`es-Nancy,
France}
\address[di]{FEMTO-ST institute, D\'{e}partement d'Informatique des
Syst\`{e}mes Complexes,\\
University of Franche-Comt\'{e}\\
16 route de Gray, 25030 Besan\c{c}on Cedex, France}
\address[tf]{FEMTO-ST institute, D\'{e}partement Temps-Fr\'{e}quence,\\
University of Technology of Belfort-Montb\'{e}liard\\
26 chemin de l'Epitaphe, 25030 Besan\c{c}on Cedex, France}

\input abstract.tex


\end{frontmatter}

\section{Introduction}
\input introduction.tex

\section{Preliminaries and notations}
\label{prelim:sec}
\input rewrite.tex

\subsection{Running example}
\label{example:sec}
\input example.tex

\section{Transformation language}
\label{package:sec}
\input rhomaple.tex

\section{Advanced features}
\label{advanced:sec}
\input advanced.tex

\section{Formal proof examples}
\label{proof:sec}
\input {gradient-input-explanations.tex}



\input encoding.tex

\section{Theoretical basis and related work}
\label{related:sec}
\input related.tex

\section{Conclusion}
\label{conclusion:sec}
\input conclusion.tex

\begin{ack}
The authors would like to thank the anonymous reviewers of a previous version
of this paper for their helpful comments, and R.
N. Dhara and B. Yang for the feedback provided by their development activity
using \ST.
\end{ack}

\bibliographystyle{elsart-harv}
\bibliography{biblioJSC}


\appendix
\section{A mathematical proof of the derivative weak convergence
property}
\label{appendix:gradient:math:proof}
\input{GradientProof/gradient.tex}

\section{A mathematical two-scale transformation of the heat equation}
\label{appendix:diffusion:math:proof}
\input{GradientProof/diffusion_equation.tex}

\section{A formal proof of the derivative weak
convergence property}
\label{appendix:gradient:maple:proof}
\begin{small}
\begin{verbatim}
# File gradient.mpl
# Contributors: Walid Belkhir, Alain Giorgetti and Michel Lenczner

# Execution (with Maple 10 or more):
#  cmaple -s -q < gradient.mpl

# Documentation: see the paper entitled
#  "A Symbolic Transformation Language and its Application to a Multiscale Method".

# 1. Library loading

new_lib_dir := "../lib":
libname := new_lib_dir, libname:
with(stmodule);
# The list of module functions is printed.

# 2. Rewrite systems loading

read `integral.mpl`:
read `Green.mpl`:
read `twoscale.mpl`:
read `convergence.mpl`:
read `lemmas.mpl`:
read `hypothesis.mpl`:

# 3. Proof

# Initial term
LHSterm := SUM(
  Integral(
    [TSTM(Omega),TSTm(Omega)],
    T(partial(u,x(i))) * v(i),
    [x,y]),
  i,Iset);

# Step 1
LHSterm := Outermost(TwoScaleAdjointInv)(LHSterm);

# Step 2
LHSterm := Outermost(ApproximationTS[1])(LHSterm);
LHSterm := evala(Expand(LHSterm));
LHSterm := IntegrationStrategy(LHSterm);
LHSterm := ConvergenceStrategy(LHSterm);

# Step 3
LHSterm := Outermost(GreenRule)(LHSterm);
LHSterm := OutermostNF(IntegralOnBoundary(v(i))(partialBound))(LHSterm);
LHSterm := IntegrationStrategy(LHSterm);

# Step 4
LHSterm := Outermost(PartialOfB)(LHSterm);
LHSterm := evala(Expand(LHSterm));
LHSterm := IntegrationStrategy(LHSterm);

# Step 5
ApproximationTSInv2Context :=
 Comp([
  [(1/Epsilon)*X_,(1/Epsilon)*X],
  Outermost(ApproximationB[2](x(j))(y(j))(j))
 ]);
LHSterm := OutermostNF(ApproximationTSInv2Context)(LHSterm);
LHSterm := Outermost(ApproximationB[1])(LHSterm);
LHSterm := evala(Expand(LHSterm));
LHSterm := Outermost(LinearityOf[TS])(LHSterm);
LHSterm := evala(Expand(LHSterm));
LHSterm := Outermost(LinearityBasic[TS])(LHSterm);
LHSterm := evala(Expand(LHSterm));
LHSterm := IntegrationStrategy(LHSterm);
LHSterm := ConvergenceStrategy(LHSterm);

# Step 6
LHSterm := Outermost(TwoScaleAdjoint)(LHSterm);
LHSterm := evala(Expand(LHSterm));
LHSterm := IntegrationStrategy(LHSterm);

# Step 7
ApproximationT2Context :=
 Comp([
  [(1/Epsilon)*X_,(1/Epsilon)*X],
  Outermost(ApproximationT[2])
 ]);
LHSterm := Outermost(ApproximationT2Context)(LHSterm);
LHSterm := Outermost(ApproximationT[1])(LHSterm);
LHSterm := evala(Expand(LHSterm));
LHSterm := IntegrationStrategy(LHSterm);
LHSterm := ConvergenceStrategy(LHSterm);
AdhocSimplify := [
  Integral(Omega_,C_*SUM(F_,J_,D_),X_),
  SUM(Integral(Omega,C*F,X),J,D)];
LHSterm := Outermost(AdhocSimplify)(LHSterm);

# Steps 8 and 9 of the mathematical proof.
# Differ from the mathematical proof: All the Green rule
# are performed.

LHSterm := STNormalizer(
 OutermostNF(
  Comp([
    OutermostNF(CondGreenRule(v(i))),
    Comp([OutermostNF(IntegralOnBoundary(v(i))(partialBound)),IntegrationStrategy])
  ])
 )
)(LHSterm);

LHSterm := OutermostNF(U0IndependentOfY)(LHSterm);
LHSterm := IntegrationStrategy(LHSterm);
LHSterm := OutermostNF(IndependentOfX)(LHSterm);

# Right-hand side term

RHSterm :=
 SUM(
   Integral(
     [TSTM(Omega),TSTm(Omega)],
     (partial(u0,x(i)) + partial(u1,y(i)))* v(i),
     [x,y]),
   i,Iset)
 + BigO(FreshIndex(),Epsilon);
RHSterm := evala(Expand(RHSterm));
RHSterm := IntegrationStrategy(RHSterm);

# ---- The left- and right-hand side terms match modulo O(epsilon).

Result := ConvergenceStrategy(LHSterm-RHSterm);
quit;
\end{verbatim}
\end{small}

%
%

\section{A formal two-scale transformation of the heat equation}
\label{appendix:diffusion:maple:proof}
\begin{small}
\begin{verbatim}
# Copyright (c) 2010-2011, University of Franche-Comt'e
# All rights reserved.
# Redistribution and use in source and binary forms, with or without
# modification, are permitted provided that the following conditions are met:
#
#     * Redistributions of source code must retain the above copyright
#       notice, this list of conditions and the following disclaimer.
#     * Redistributions in binary form must reproduce the above copyright
#       notice, this list of conditions and the following disclaimer in the
#       documentation and/or other materials provided with the distribution.
#     * Neither the name of the University of Franche-Comt'e nor the
#       names of its contributors may be used to endorse or promote products
#       derived from this software without specific prior written permission.
#
# THIS SOFTWARE IS PROVIDED BY THE CONTRIBUTORS ``AS IS'' AND ANY
# EXPRESS OR IMPLIED WARRANTIES, INCLUDING, BUT NOT LIMITED TO, THE IMPLIED
# WARRANTIES OF MERCHANTABILITY AND FITNESS FOR A PARTICULAR PURPOSE ARE
# DISCLAIMED. IN NO EVENT SHALL THE CONTRIBUTORS BE LIABLE FOR ANY
# DIRECT, INDIRECT, INCIDENTAL, SPECIAL, EXEMPLARY, OR CONSEQUENTIAL DAMAGES
# (INCLUDING, BUT NOT LIMITED TO, PROCUREMENT OF SUBSTITUTE GOODS OR SERVICES;
# LOSS OF USE, DATA, OR PROFITS; OR BUSINESS INTERRUPTION) HOWEVER CAUSED AND
# ON ANY THEORY OF LIABILITY, WHETHER IN CONTRACT, STRICT LIABILITY, OR TORT
# (INCLUDING NEGLIGENCE OR OTHERWISE) ARISING IN ANY WAY OUT OF THE USE OF THIS
# SOFTWARE, EVEN IF ADVISED OF THE POSSIBILITY OF SUCH DAMAGE.

# File heat.mpl
# Contributors: Walid Belkhir, Alain Giorgetti and Michel Lenczner

# Execution (with Maple 10 or more):
#  cmaple -s -q < heat.mpl

# Documentation: see the paper entitled
#  "A Symbolic Transformation Language and its Application to a Multiscale Method".

# 1. Library loading

new_lib_dir := "../lib":
libname := new_lib_dir, libname:
with(stmodule);

# 2. Rewrite systems loading

read `integral.mpl`:
read `Green.mpl`:
read `twoscale.mpl`:
read `convergence.mpl`:
read `lemmas.mpl`:
read `hypothesis.mpl`:

# 3. Lemmas

GradientApprox := [
  SUM(
    Integral(
      [TSTM(Omega), TSTm(Omega)],
      a0*T(partial(U_,X1_))*V_,
      [X2_, Y_]),
    Iset_,K_),
  DelayEval -> SUM(
    Integral(
      [TSTM(Omega), TSTm(Omega)],
      a0*(partial(u0,X1)+partial(u1,Y(Iset)))*V,
      [X2, Y]),
    Iset,K)+BigO(FreshIndex(),Epsilon)];
  
# 4. Proof
  
# Initial term. The equality is replaced by a difference
# to get more simplifications.
EqTerm := SUM(
  Integral(
    Omega,
    a * partial(u,x(i)) * partial(v,x(i)),
    x),
  i,Iset) - Integral(Omega,f*v,x);

# Step 1
TestFunChoice[1] := [v,B(v0+Epsilon*v1)];
EqTerm := Outermost(TestFunChoice[1])(EqTerm);

# Step 2
EqTerm := Outermost(PartialOfB)(EqTerm);
EqTerm := Outermost(LinearityOf[partial])(EqTerm);
EqTerm := Outermost(Linearity2Of[partial])(EqTerm);
EqTerm := Outermost(LinearityOf[B])(EqTerm);
EqTerm := Outermost(Linearity2Of[B])(EqTerm);
EqTerm := evala(Expand(EqTerm));
EqTerm := Outermost(V0IndependentOfY)(EqTerm);
EqTerm := Outermost(V1IndependentOfX)(EqTerm);
EqTerm := Outermost(LinearityOf[B])(EqTerm);

# Extra ad hoc factorization:
factor1 := [a*partial(u,x(i))*X_+a*partial(u,x(i))*Y_,a*partial(u,x(i))*(X+Y)];
EqTerm := Outermost(factor1)(EqTerm);

# Ad hoc inverse linearity rule.
factor2 := [B(X_)+B(Y_), B(X+Y)];
EqTerm := Outermost(factor2)(EqTerm);

# Step 3
EqTerm := Outermost(ApproximationB[1])(EqTerm);
EqTerm := ConvergenceStrategy(EqTerm);

# Step 4
EqTerm := Outermost(TwoScaleAdjoint)(EqTerm);

# Step 5
EqTerm := Outermost(ApproximationOfTHypo)(EqTerm);
EqTerm := Outermost(ApproximationT[3])(EqTerm);
EqTerm:=ConvergenceStrategy(EqTerm);

# Step 6
EqTerm := Outermost(LinearityOf[SUM])(EqTerm);
EqTerm := Outermost(GradientApprox)(EqTerm);
EqTerm:=ConvergenceStrategy(EqTerm);

quit;
\end{verbatim}
\end{small}

\section{Rule and transformation files}
\label{appendix:rules}


\subsection{Domain integral and indefinite sum rules}
\label{IntegralRules:sec} 

\begin{small}
\begin{verbatim}
# File:    integral.mpl.
# Content: Rules for integral and indefinite summation properties.

LinearityOf[Integral]  := Linearity(2,x->y->x+y,Integral(Omega_,_,Z_));
Linearity2Of[Integral] := [Integral(Omega_,0,X_),0];

LinearityOf[partial]  := Linearity(1,x->y->x+y,partial(_,B_));
Linearity2Of[partial] := Linearity2(1,partial(X_,Y_),Epsilon );

LinearityLambdaOf[Integral] := [
  X_*Integral(Omega_,Y_,Z_),
  Integral(Omega,X*Y,Z)];

LinearityOf[SUM] := Linearity(1,x->y->x+y,SUM(_,J_,K_));

SumZero := [SUM(0,X_,J_),0];

SumInteger := [
  SUM(A_*B_,J_,D_),
  A*(SUM(B,J,D)),
  DelayEval -> whattype(A)=integer];

# To group two sums into a single one:
SumFactor := [
  SUM(E_,I_,D_) + SUM(F_,I_,D_),
  DelayEval -> SUM(factor(E+F),I,D)];

# WARNING: Do not include the following rule in the same strategy as a
# factorization rule. May loop forever!

SumDistrib := [
  A_*(SUM(B_,J_,D_)),
  SUM(A*B,J,D)];

IntegralSumExchange := [
  Integral(M_,SUM(A_,J_,D_),B_),
  SUM(Integral(M,A,B),J,D)];

IntegrationStrategy :=
  STNormalizer(
    FailAsIdentity(
     LeftChoice([
       Outermost(LinearityOf[Integral]),
       Outermost(Linearity2Of[Integral]),
       Outermost(LinearityOf[SUM]),
       Outermost(SumZero),
       Outermost(SumInteger),
       Outermost([partial(0,X_),0])
     ])
    )
  );
\end{verbatim}
\end{small}

\subsection{Green rules}
\label{GreenRule:maple} 

\begin{small}
\begin{verbatim}
# File:    Green.mpl.
# Content: Green rules.

GreenRule := [
  Integral(Omega_,V_*partial(U_,X_),Y_),
  Integral(partialBound([Omega,X]),U*V*Eta(X),s(X))
   -Integral(Omega,U* partial(V,X),Y)];

CondGreenRule := patt -> [
  Integral(Omega_,V_*partial(U_,X_),Y_),
  Integral(partialBound([Omega,X]),U*V*Eta(X),s(X))
   -Integral(Omega,U* partial(V,X),Y),
  DelayEval-> has(U,patt)];
\end{verbatim}
\end{small}

\subsection{Convergence rules}
\label{convergence:rules:sec} 

\begin{small}
\begin{verbatim}
# File:    convergence.mpl.
# Content: Convergence theory and strategy.

OEpsilonSum := [
  BigO(I_,Epsilon) + BigO(J_,Epsilon),
  BigO(FreshIndex(),Epsilon)];

OEpsilonSumContext := [
  Y_+ BigO(I_,Epsilon) + BigO(J_,Epsilon),
  DelayEval -> BigO(FreshIndex(),Epsilon)+Y];

OEpsilonSUM := [
  SUM(BigO(I_,Epsilon),J_,D_),
  DelayEval -> BigO(FreshIndex(),Epsilon)];

OEpsilonSUMContext := [
  SUM(BigO(I_,Epsilon),J_,D_)+Z_,
  DelayEval -> BigO(FreshIndex(),Epsilon)+Z_];

OEpsilonIntegral := [
  Integral(Omega_, BigO(I_,Epsilon), X_),
  DelayEval -> BigO(FreshIndex(),Epsilon)];

# Controlled linearity
OEpsilonIntegralContext := [
  Integral(Omega_, BigO(I_,Epsilon)+E_, X_),
  DelayEval -> Integral(Omega, E, X)
    + BigO(FreshIndex(),Epsilon)];

# Controlled expansion
OEpsilonExpand := [
  A_ * (E_ + BigO(I_,Epsilon)),
  DelayEval -> A * E + A * BigO(I,Epsilon)];

# Only if A_ does not depend on Epsilon.
OEpsilonConst := [
  A_ * BigO(I_,Epsilon),
  DelayEval -> BigO(FreshIndex(),Epsilon)];

# Only if A_ does not depend on Epsilon.
EpsilonConst:= [
  A_ * Epsilon,
  DelayEval -> BigO(FreshIndex(),Epsilon)];

ConvergenceStrategy :=
  STNormalizer(
    FailAsIdentity(
      LeftChoice([
        Outermost(OEpsilonSum),
        Outermost(OEpsilonSumContext),
        Outermost(OEpsilonSUM),
        Outermost(OEpsilonSUMContext),
        Outermost(OEpsilonExpand),
        Outermost(OEpsilonIntegral),
        Outermost(OEpsilonIntegralContext),
        Outermost(OEpsilonConst),
        OutermostNF(EpsilonConst)
      ])
    ) 
  );
\end{verbatim}
\end{small}

\subsection{Two-scale method rules}
\label{two-scale:mpl:annex} 

\begin{small}
\begin{verbatim}

LinearityOf[T] := Linearity(1,x->y->x+y,T(_));
LinearityBasic[TS] := Linearity(1,x->y->x+y,TS(_));

TwoScaleMult := [T(X_*Y_),T(X)*T(Y)];

TwoScaleAdjoint := [
  Integral(Omega_, TS(V_)*W_ , X_),
  Integral([TSTM(Omega) , TSTm(Omega)], V*T(W), [X, y])];

TwoScaleAdjointInv := [
  Integral([TSTM(Omega_) , TSTm(Omega_)], T(W_)*V_, [X_, Z_]),
  Integral(Omega, W*TS(V) , X)];

PartialOfB := [
  partial(B(V_),x_(I_)),
  B(partial(V,x(I))) + 1 / Epsilon*B(partial(V,y(I)))];

ApproximationTS := [
  [TS(V_),DelayEval -> B(V)+ BigO(FreshIndex(),Epsilon)],
  [TS(V_), DelayEval -> TS(V) + Epsilon * TS(y*partial(V,x))
              + Epsilon * BigO(FreshIndex(),Epsilon)]
];

ApproximationB := [
  [B(V_), DelayEval -> TS(V)+BigO(FreshIndex(),Epsilon)],
  x -> y -> j -> [
     B(V_),
     TS(V+Epsilon*SUM(y*partial(V,x),j,J)) 
      + Epsilon*BigO(FreshIndex(),Epsilon)
  ]
];

LinearityBasic[B] := LeftChoice([
  [B(0),0],
  Linearity(1,x->y->x+y,B(_))
]);

LinearityOf[B] :=
  STNormalizer(
    OutermostNF(
      IdentityAsFail(LinearityBasic[B])
    )
  );

Linearity2Of[B] := [B(Epsilon*F_),Epsilon*B(F)];

# The following rule has to be generalized. 
# We need to include, propagate and check
# the type of expressions and operators.
LinearityOf[TS] := [A_*(1/Epsilon)*TS(V_),A*TS(1/Epsilon*V)];
\end{verbatim}
\end{small}


\subsection{Lemmas}
\label{lemmas:maple:appendix} 

\begin{small}
\begin{verbatim}
# File:    lemmas.mpl.
# Content: Lemmas to prove separately.

# Zero- and first- order approximations of T.
ApproximationT := [
  [T(u), DelayEval -> u0 + BigO(FreshIndex(),Epsilon)],
  [T(u), u0 + Epsilon*u1 + Epsilon*SUM(y(j)*partial(u0,x(j)),j,J)
         + Epsilon*BigO('FreshIndex()',Epsilon)],
  [T(f), DelayEval -> f0 + BigO(FreshIndex(),Epsilon)]
];

IndependentOfX := [
  partial(y(I_)*A_,x(I_)),
  y(I)*partial(A,x(I))];

ApproximationOfTHypo := [
  T(a*partial(F_,X_)),
  DelayEval-> a0*T(partial(F,X)) + BigO(FreshIndex(),Epsilon) ];

# Proved in u0IndependentOfYLemma.mpl.
U0IndependentOfY := [
  partial(Const_*u0,y(i)),
  0
];
\end{verbatim}
\end{small}

\subsection{Hypotheses}
\label{hypotheses:maple:appendix} 
\begin{small}
\begin{verbatim}
# File:    hypothesis.mpl.
# Content: Proof-dependent rules.

# Behavior of v on the boundary
OnBoundary[v] := [v(i),0];

BoundaryContext := B -> [ 
  Integral(Omega_,F_,X_),
  Integral(Omega,F,X),
  DelayEval-> has(Omega,B)];

# Integral of <w> is 0 on the boundary of <Domain> 
IntegralOnBoundary:= w -> Domain ->
  Comp([
    BoundaryContext(Domain),
    OutermostNF([w,0]),
    OutermostNF([B(0),0])
  ]);

V0IndependentOfY := [
  partial(v0,y(i)),
  0
];

V1IndependentOfX := [
  partial(v1,x(i)),
  0
];

V0IndependentOfYIntegral := [
  Integral(Omega_,v0*E_,D_),
  v0*Integral(Omega,E,D)
];

PartialXjU0IndependentOfY := [
  Integral(Omega_,partial(u0,x(j))*E_,D_),
  partial(u0,x(j))*Integral(Omega,E,D)
];

PartialXkV0IndependentOfY := [
  Integral(Omega_,partial(v0,x(k))*E_,D_),
  partial(v0,x(k))*Integral(Omega,E,D)
];
\end{verbatim}
\end{small}



\end{document}

%% file: abstract.tex
\begin{abstract}
The context of this work is the design of a software, called \emph{MEMSALab},
dedicated to the automatic derivation of multiscale models of arrays of micro-
and nanosystems. In this domain a model is a partial differential equation.
Multiscale methods approximate it by another partial differential equation which
can be numerically simulated in a reasonable time.    The challenge consists in 
 taking into account a wide range of geometries combining thin and  periodic
structures with the possibility of multiple nested scales.

In this paper we present a transformation language that will
make the development of MEMSALab more feasible. It is proposed as a
Maple$^{\mathsf{TM}}$ package for rule-based programming, rewriting strategies and their combination
 with standard Maple$^{\mathsf{TM}}$ code.
We illustrate the practical interest of this language by using it to encode two
examples of multiscale derivations, namely the two-scale limit of 
the derivative operator and the two-scale  model  of the stationary heat equation.
\end{abstract}


%% file: introduction.tex
The context of this work is the design of microsystem array architectures,
including microcantilevers, micromirrors, droplet ejectors, micromembranes,
microresistors, etc., to cite only a few. A model for such arrays is a Partial
Differential Equation (PDE). The numerical simulation of whole arrays based on
classical methods like the Finite Element Method (FEM) is prohibitive for
today's computers (at least in a time compatible with the time scale of a
designer). The calculation of a reasonably complex cell of a three-dimensional
microsystem requires at least 1000 degrees of freedom which lead to at least 10
000 000 degrees of freedom for a 100 $\times$ 100 array. Fortunately there is a
solution consisting in approximating the model by a multiscale method. The
resulting approximated model is again a PDE. It can be rigorously derived from
the exact one through a sequence of mathematical transformations, but these
transformations differ for each case.

We are currently developing a software, called MEMSALab, for ``MEMS Arrays
Laboratory'', dedicated to multiscale and multiphysics modeling of arrays of
micro- and nanosystems. Unlike traditional software that is based on models
built once and for all, MEMSALab is a software that constructs models.  The
challenge consists in   taking into account a wide range of geometries combining
thin and periodic structures with the possibility of multiple nested scales. One
should also consider PDEs representing multiphysics systems with high contrast
in equation coefficients.

Simulation software available in the market offers specialized tools for
 large arrays of micro- and nanosystems, but the construction of new models
 raises many problems. Firstly the time required for a new design varies from
 some weeks for a specialist to several months for a beginner. Secondly the
 mathematical machinery is too sophisticated to be manually applied to complex
 systems. Finally the resulting models require specific
 numerical simulation methods, that have to be implemented case by case.

The software MEMSALab we design aims at addressing these problems. It is based
on multiscale models, especially on those derived by asymptotic methods.  Such
asymptotic models are derived from a system of PDEs when taking into account
that at least one parameter is very small, such as thickness for a thin
structure or the small ratio of a cell size to the global size for a periodic
structure. The resulting models are other systems of PDEs, obtained by taking
the mathematical limits of the nominal models, in a well-suited  sense, when the
small parameters tend toward zero. This approach provides a reasonably good
approximation. It also offers the advantages and factors of reliability to be
rigorous and systematic. The resulting PDEs can be implemented in a simulation
software such as the finite element based simulator COMSOL (Multiphysics Finite
Element Analysis Software, official site \url{http://www.comsol.com}), and
simulations turn to be fast as needed.

The literature in this field is vast and a large number of techniques have been
developed for a large variety of geometric features and physical phenomena.
However, none of them have been implemented in a systematical approach to render
it available to engineers as a design tool. In fact, each published paper focus
on a special case regarding geometry or physics, and very few works are
considering a general picture. By contrast our software will treat the problem
of systematic implementation of asymptotic methods by implementing the
construction of models rather than the models themselves. This approach will
cover many situations from a small number of bricks. It combines mathematical
and computer science tools. The mathematical tool is the two-scale transform
originally introduced in \citep{Len97, Cas00, CioDam02} to model periodic and
thin structures, and also referred as the unfolding method. We have extended its
domain of application to cover in the same time homogenization of periodic
media, see for instance \cite{BenLio}, and methods of asymptotic analysis for
thin domains, see \cite{Cia}. The computer science tools include term rewriting,
$\lambda$-calculus, and type systems~\citep{rhoCalIGLP-I+II-2001, MP_RhoLog04,
CKL01, Geuvers:2009:ITT}. The software is written in the symbolic computation
language Maple$^{\mathsf{TM}}$.

Compared to other techniques, our multiscale method requires more modular
calculations, avoids any non-constructive proof and intensively relies
on equational reasoning. The classical way to automate equational reasoning
is to consider mathematical equalities as rewrite rules. 
 The rewrite rule $\rrule{t}{u}$ orients the equality $t=u$ from left to
 right and states that every occurrence of an instance of $t$ can be replaced
 with the corresponding instance of $u$. Consequently symbolic computation with equalities is reduced
  to a series of term rewritings. Algebraic computation and term rewriting are two research domains with
strong similarities. Both are separately well-studied but there are only few
works about the combination of algebraic computation and term rewriting
~\citep{Fevre:1998:CAC,Bundgen95combiningcomputer}.
  Term rewriting provides a theoretical and computational framework which is very
useful to express, study and analyze a wide range of complex systems. It
is characterized  by a repeated transformation of  data objects such as  words,
terms or graphs. Transformations are described by a combination of rules which
specify how to transform an object into another one in the presence of a specific
pattern. Rules can have further conditions and can
 be combined by specifying strategies.  The latter control the order and the way
 the rules are applied.
Term rewriting is  used in semantics in order to describe the meaning of
programming languages as well as in program transformations. It is used to
perform symbolic computations like in Mathematica, and also to perform
automated reasoning. It is central in systems where the notion of \emph{rule}
is explicit such as expert systems, algebraic specifications, etc.

The  computer algebra system Maple$^{\mathsf{\tiny TM}}$  is widely used in the
symbolic computation community. It is also used  by members of our project for a
prototypal implementation of MEMSALab.
Maple$^{\mathsf{TM}}$ is a suitable language for combining function-based
 and rule-based symbolic
transformations.  Unfortunately, it is only equipped with a limited rewrite
kernel, namely the function \texttt{apply\-rule(rule,expr)}. The main drawback
of \texttt{applyrule} is that it iterates the application of the rule everywhere
in the given expression until no matching sub-expression remains. Therefore
there is a lack of flexibility: the user cannot express how and where rules must
be applied. The other problem is that the Maple$^{\mathsf{TM}}$ matching
 function \texttt{patmatch(expr,pattern,sub)}
has two main drawbacks. Firstly,  when the user defines new operators, the
matching must be done modulo the properties  of these operators, e.g.
associativity and commutativity. Actually, \texttt{patmatch} does not provide
this feature. Secondly the matching function computes just one solution (i.e. a
substitution) of the matching problem of \texttt{pattern} with \texttt{expr},
whereas both supporting associative-commutative operators and implementing
conditional rewrite rules require a matching that returns
all the possible solutions.

\subsection*{Contributions} 

In this work we present a transformation language named \ST (for ``symbolic
transformation''). It extends Maple$^{\mathsf{TM}}$ with conditional rewriting,
strategies, and pattern-matching modulo associativity and commutativity. All the
ingredients of term rewriting with strategies are made explicit. In particular
terms, patterns, rules, strategies, pattern matching and application of rules
and strategies to terms are represented by Maple$^{\mathsf{TM}}$ expressions.
Rewrite rules and rewriting strategies are deterministic functions that raise an
exception when they are not applicable. Such functions can be combined with
Maple$^{\mathsf{TM}}$ functions. We illustrate the practical interest of this
transformation language by using it to write two
formal derivations for the MEMSALab software. The first one states the 
weak two-scale limit  of the derivative operator. The second one computes the
two-scale model of the stationary heat equation.

\subsection*{Related work} 

Our  transformation language is an adaptation for Maple$^{\mathsf{TM}}$ of popular strategy
languages such as $\rho$-log~(\cite{MP_RhoLog04}) or
\textsc{Tom}~(\cite{BBK_Tom07}). A conceptual difference
  with  \textsc{Tom} is that the latter  extends a host language with an
  additive syntax, whereas our transformation language smoothly integrates with standard
Maple$^{\mathsf{TM}}$ functions. The first work that have 
considered term rewriting from a functional point
of view is Elan, see  \citep{BKK+99}, within a non-deterministic framework.
Our transformation language is comparable  with the  deterministic fragment of
 the  \emph{rewriting calculus}  \citep{rhoCalIGLP-I+II-2001}.

\subsection*{Paper outline} 

Section \ref{prelim:sec} introduces the two-scale transform and necessary term
rewriting concepts and notations. Section \ref{package:sec} defines the
transformation language \ST in a detailed way and illustrates transformations by
several examples. Section \ref{advanced:sec} presents more advanced features of
\ST: the combination of term rewriting with procedural programming, a delayed
procedure evaluation mechanism, pattern-matching modulo associativity and
commutativity, and conditional rewriting. Section~\ref{proof:sec} shows the
transformation language at work on two realistic examples of multiscale
derivations. Section~\ref{related:sec} presents the theoretical bases of \ST and
compares the present work with related ones. Section~\ref{conclusion:sec} draws
conclusions.

A mathematical proof of the weak two-scale limit property of the derivative
operator   is reproduced  in Appendix \ref{appendix:gradient:math:proof}. Its
formal proof with \ST is reproduced in Appendix
\ref{appendix:gradient:maple:proof}. A mathematical derivation of the 
two-scale  model of the stationary heat equation is reproduced in
Appendix~\ref{appendix:diffusion:math:proof}. Its formal counterpart
with \ST is given in Appendix \ref{appendix:diffusion:maple:proof}.
Rules and transformations corresponding to mathematical properties for these
formal proofs are reproduced in Appendix~\ref{appendix:rules}.


%% file: rewrite.tex
\subsection{Term, substitution, matching and rewriting}

Term rewriting systems (in the classical sense) are defined by specifying a set
of terms and a set of rewrite rules. Rewrite rules are applied to \emph{reduce}
terms.  In general the process of reduction continues until no more rules can be
applied, or forever in the case of non-terminating systems. A term which cannot
be reduced to another term is called a \emph{normal form}. If there is always a
unique normal form then the system is said to be \emph{confluent}.

Let  $\mathcal{F}$ be a countable set of function symbols, each
symbol having a fixed arity. Let $\mathcal{X}$ be a countable set of
variables. The set of terms, denoted by
$\mathcal{T(F,\mathcal{X})}$, is inductively
 defined as the smallest set containing the elements of $\mathcal{X}$ and $f(t_1, \ldots, t_n)$,
 for any symbol $f$ with arity $n$ in $\mathcal{F}$ and any terms $t_1$, \ldots,
  $t_n$ in $\mathcal{T(F,\mathcal{X})}$.
When $n=0$ the symbol $f$ is called a \emph{constant} and the corresponding
term is denoted $f$ instead of $f()$.

The set of variables occuring in a term $t$ is denoted by
$\free{t}$. If $\free{t}=\emptyset$ then $t$ is said to be
\emph{ground}. A \emph{substitution} is a function $\sigma$
 from $\mathcal{X}$ to $\mathcal{T(F,X)}$ such that $\sigma(x) \neq x$
 only for finitely many variables $x$ in $X$. If $x_1$, \ldots, $x_n$ 
 are these variables, then $\sigma$
is denoted by $\{x_1\mapsto t_1,\ldots,x_n\mapsto t_n\}$, where $t_i=\sigma(x_i)$ ($1 \leq i \leq n$). If $t$
is a term then  $\sigma(t)$ is the term that results from the
application of $\sigma$ to $t$.

A \emph{rewrite rule} is a pair $(l,r)$ of terms $l$ and $r$
 in $\mathcal{T(F,\mathcal{X})}$ s.t. $\free{r} \subseteq
\free{l}$. This pair is usually written $\rrule{l}{r}$.

\begin{defn}\label{matching:def}
For two terms $t$ and $t'$ in $\mathcal{T(F,\mathcal{X})}$
the problem of finding substitutions $\sigma$ such that $\sigma(t')=t$ is called 
a \emph{matching problem} and is denoted $\match{t'}{t}$.  A substitution
$\sigma$ such that $\sigma(t')=t$ is called a \emph{solution} of this matching
problem.
The \emph{application} of a rewrite rule $\rrule{l}{r}$ to a term
$t$, denoted by $[\rrule{l}{r}](t)$, is defined
by $\sigma(r)$ where $\sigma$ is any solution of the matching problem
$\match{l}{t}$. It is undefined when this matching problem has no solution.
\end{defn}

\begin{exmp}
Let $f$, $g$ , $+$ and $*$ (resp. $a$ and $b$) be
function symbols of arity $2$ (resp. $0$) in $\mathcal{F}$. Let $x$ and $y$ be
variables in $\mathcal{X}$.
\begin{itemize}
\item The application $[\rrule{x}{b}](f(a,b))$ of the rule $\rrule{x}{b}$ to
the term $f(a,b)$ is the term $b$. The substitution $\sigma=\{x\mapsto
f(a,b)\}$ is an obvious solution of the matching problem $\match{x}{f(a,b)}$.
\item The application $[\rrule{f(x,a)}{g(a,x)}](f(b,a))$ of the rule 
$\rrule{f(x,a)}{g(a,x)}$ to the term $f(b,a)$ is the term $g(a,b)$, with
 the substitution $\sigma=\{ x\mapsto b\}$.
\item The application $[\rrule{a}{b}](a)$ of the rule $\rrule{a}{b}$ to the term $a$ is the term $b$, with the empty
substitution $\sigma=\{\}$.
\end{itemize}
\end{exmp}
It is worth  mentioning that there is a difference between the
notions of variables  and constants in the mathematical sense and in
 the sense of term rewriting when mathematical expressions are viewed as terms.
Let us give a concrete example. Consider the mathematical expression
$\int_{\Omega}f(x) dx$ and its representation by
the term \texttt{Integral(Omega,} \texttt{f(x),} \texttt{x)}.
In this term, \texttt{Omega} and the mathematical variable \texttt{x} are
viewed as function symbols  of arity $0$ (i.e. constants), \texttt{f} is
viewed as a function symbol of arity $1$, and \texttt{Integral} as a function
symbol of arity $3$. Further clarifications on this topic can be found in Section
\ref{package:sec}.


%% file: example.tex
We illustrate the transformation language with the homogenization
problem of the stationary heat equation. We consider a stationary
distribution of temperature in a region $\Omega \subset
\mathbb{R}^{n}$ (where $n\in \{1,2,3\}$) with an internal heat
source and with imposed vanishing
temperature along the boundary. The diffusion coefficient $a^{\varepsilon }:%
\mathbb{R}^{n}\rightarrow \mathbb{R}$ is assumed to be periodic on
$\Omega $ in the $n$ directions, with a small period $\varepsilon $.
In other words,
there is a function $a:\mathbb{R}^{n}\rightarrow \mathbb{R}$ which is $%
(0,1)^{n}$-periodic and such that $a^{\varepsilon
}(x)=a(x/\varepsilon )$ for $x\in \Omega $. With a view to derive
the so-called \textit{homogenized model}, the parameter $\varepsilon
$ is considered as small, and we are
interested in finding an approximation of the stationary heat equation when $%
\varepsilon $ decreases to zero. In this mathematical asymptotic
process,
the distributed internal heat source $f^{\varepsilon }:\mathbb{R}%
^{n}\rightarrow \mathbb{R}$ can be considered as depending on $\varepsilon $%
, and the temperature distribution $u^{\varepsilon }:\mathbb{R}%
^{n}\rightarrow \mathbb{R}$, vanishing on the boundary $\partial
\Omega $ of $\Omega $, is the unique solution to the stationary heat
equation
\begin{equation}
\sum_{i=1}^{n}\int_{\Omega }a^{\varepsilon
}\partial_{x_i}u^{\varepsilon }\,\partial_{x_i}v\text{ }
dx=\int_{\Omega }f^{\varepsilon }v\text{ }dx
\label{model:diffusion:variational:eq}
\end{equation}%
written in its variational form, as explained in ~\citep{DauLio}. Here, $%
v:\Omega \rightarrow \mathbb{R}$ is any \emph{test function}, i.e. a
sufficiently regular function vanishing on $\partial \Omega.$

To conduct the asymptotic process $\varepsilon \rightarrow 0$ while
keeping as much information as possible in the solution
$u^{\varepsilon }$ at the
small scale, the region $\Omega $ is unfolded into the Cartesian product $%
\widetilde{\Omega }\times Y$ with $Y=(0,1)^{n}$ through the
so-called \textit{two-scale transform} or \textit{unfolding}, where
$\widetilde{\Omega }\subset \mathbb{R}^{n}$ is called the
\emph{macroscopic domain} associated
to $\Omega $, and $Y$ is called the \emph{microscopic domain} associated to $%
\Omega $. This sort of change of variables is applied to the sequence $%
u^{\varepsilon }:\Omega \rightarrow \mathbb{R}$ which yields another
sequence of functions $Tu^{\varepsilon }:\widetilde{\Omega }\times
Y\rightarrow \mathbb{R}$. We can show that the latter converges to a limit $%
u^{0}(x,y)$, so we say that the sequence $u^{\varepsilon }$ is \textit{%
two-scale convergent }to $u^{0}.$ Similarly, we show that
$a^{\varepsilon }$, $f^{\varepsilon }$ and $\partial _{x_{i}}u^{\varepsilon }$ 
 are two-scale convergent towards some limits
$a^{0},$ $f^{0}$ and $\partial _{x_{i}}u^{0}+\partial
_{y_{i}}u^{1},$ and in the same time that $u^{0}$ is independent of
the so-called \textit{microscopic variable }$y$, it is vanishing on
$\partial \widetilde{\Omega }$ and $u^{1}$ is $Y$-periodic, a
concept that is explained later. Then, applying the two-scale
transform in the variational formulation
(\ref{model:diffusion:variational:eq}) of the heat equation, passing
to the limit $\varepsilon \rightarrow 0$, we arrive
to the \textit{two-scale model} satisfied by the pair $(u^{0},u^{1})$,%
\begin{equation}
\sum_{i=1}^{n}\int_{\widetilde{\Omega }\times Y}a^{0}\,(\partial
_{x_{i}}u^{0}+\partial _{y_{i}}u^{1})\;(\partial
_{x_{i}}v^{0}+\partial _{y_{i}}v^{1})\;dxdy=\int_{\widetilde{\Omega
}\times Y}f^{0}\,v^{0}\;dxdy, \label{model:diffusion:two-scale:eq}
\end{equation}%
where the pair $(v^{0}(x),v^{1}(x,y))$ is sufficiently regular,
$v^{0}$ is vanishing on $\partial \widetilde{\Omega }$ and $v^{1}$
is $Y$-periodic$.$ The last step consists in showing that $u^{1}$ is
a function of $u^{0}$
which allows its elimination and yields the \textit{homogenized model}%
\footnote{%
We mention that we use the minimal scope convention of the
derivative
operator, i.e. $\partial _{x_{i}}fg$ means $(\partial _{x_{i}}f)g$.}%
\begin{equation}
\sum_{j,k=1}^{n}\int_{\widetilde{\Omega }}a_{jk}^{H}\partial
_{x_{j}}u^{0}\,\partial _{x_{k}}v^{0}dx=\int_{\widetilde{\Omega }%
}f^{H}v^{0}dx  \label{model:diffusion:approx:eq}
\end{equation}%
for all test functions $v^{0}$. Here $a^{H}$ is the $n\times n$
matrix of \emph{effective diffusion coefficients}, and $f^{H}$ is
the \emph{effective
heat source}. We observe that in this example the macroscopic domain $%
\widetilde{\Omega }$ is identical to $\Omega $. However, we prefer
to distinguish them for sake of generalization, e.g. to thin
domains.

The statement of the approximate model
(\ref{model:diffusion:approx:eq}) derived with the two-scale
transform was announced in \citep{Len97}. Then
several proofs have been published, in \citep{LenSen99}, \citep{Cas00}, %
\citep{CioDam02}, \citep{LenSmi07}, and \citep{CioDam08}. Here we
follow the proof in \citep{LenSmi07} where an effort has been made to
formulate proofs in an algebraic way and to avoid any abstract
reasoning in the sequences of formal transformations. The steps in
this formal method are rigorously specified at a high level of
generality, that make them independent of the domain geometry and
applicable to other equations.

Before entering in more details, we introduce some notations and
definitions. For any region $\Theta \subset \mathbb{R}^{n},$
$L^{2}(\Theta )$ denotes the set of square integrable functions on
$\Theta ,$ that is the set of functions $v:\Theta \rightarrow
\mathbb{R}$ with a bounded $L^{2}(\Theta )
$-norm%
\begin{equation*}
||v||_{L^{2}(\Theta )}=(\int_{\Theta }v^{2}(x)\text{ }dx)^{1/2}.
\end{equation*}%
Then a sequence $u^{\varepsilon }\in L^{2}(\Theta )$ is said to be
convergent (or strongly convergent) in $L^{2}(\Theta )$ towards a limit $%
u^{0}$ if $||u^{\varepsilon }-u^{0}||_{L^{2}(\Theta )}=O(\varepsilon
)$
where $O(\varepsilon )$ is the Landau notation representing any sequence of $%
\varepsilon $ tending to zero when $\varepsilon \rightarrow 0.$
Another concept of convergence used for asymptotic models is the
concept of weak convergence. A sequence $u^{\varepsilon }\in
L^{2}(\Theta )$ is said to be weakly convergent in $L^{2}(\Theta )$
towards a limit $u^{0}$ if
\begin{equation*}
\int_{\Theta }(u^{\varepsilon }-u^{0})v\text{ }dx=O(\varepsilon
)\text{ for any }v\in L^{2}(\Theta ).
\end{equation*}%
A by product of this definition is that a sequence $u^{\varepsilon
}\in L^{2}(\Omega )$ is said to be \textit{two-scale weakly
convergent} towards a limit $u^{0}\in L^{2}(\widetilde{\Omega
}\times Y)$ if $Tu^{\varepsilon }$
is weakly convergent towards $u^{0}$ in $L^{2}(\widetilde{\Omega }\times Y)$%
. Finally, a function $v:Y\rightarrow \mathbb{R}$ is said to be
$Y$-periodic if $v(y^{+})=v(y^{-})$ for any pair of opposite points
$y^{+}$ and $y^{-}$ of the boundary of $Y$.

Now we specify the assumptions needed to build the model (\ref%
{model:diffusion:approx:eq}) as the limit of (\ref%
{model:diffusion:variational:eq}). The heat source $f^{\varepsilon
}$ is assumed to be uniformly bounded in the $L^{2}(\Omega )$-norm,
that is, there
exists a constant $C$ independent of $\varepsilon $ such that%
\begin{equation}
||f^{\varepsilon }||_{L^{2}(\Omega )}\leq C.
\label{model:diffusion:variational:hyp}
\end{equation}%
Then, the proof is divided into five parts. (i) We establish that
the fields of temperature distribution $u^{\varepsilon }$ and its
derivatives are
uniformly bounded in the $L^{2}(\Omega )$-norm,%
\begin{equation}
||u^{\varepsilon }||_{L^{2}(\Omega )}\leq C\text{ and }||\partial
_{x_{i}}u^{\varepsilon }||_{L^{2}(\Omega )}\leq C\text{ for
}i=1,...,n. \label{model:diffusion:variational:norm_estimates}
\end{equation}%
Given these results, it is assumed that $u^{\varepsilon }$ has an
asymptotic expansion on the form
\begin{equation}
Tu^{\varepsilon }=u^{0}+\varepsilon \widetilde{u}^{1}+\varepsilon
O(\varepsilon ),  \label{model:diffusion:variational:hyp1}
\end{equation}%
where here $O(\varepsilon )$ denotes a function that tends to zero
weakly in $L^{2}(\widetilde{\Omega }\times Y)$. We observe that this
assumption is not necessary to get the desired result, however, we
find that on the one hand it is not very strong once the a priori
estimates are known and on the other hand it allows the proof to be
entirely computational, i.e. without steps of
abstract reasoning. (ii) The next step consists in deducing of (\ref%
{model:diffusion:variational:norm_estimates}) that the two-scale weak limit $%
u^{0}$ of $Tu^{\varepsilon }$ is independent of $y$%
\begin{equation}
\partial _{y_{i}}u^{0}=0\text{ for all }i.  \label{gradient:cond:u0}
\end{equation}%
(iii) It comes to show that there exists $u^{1}(x,y)$ such that%
\begin{equation}
\partial _{x_{i}}u^{\varepsilon }\text{ is weakly two-scale convergent
towards }\partial _{x_{i}}u^{0}+\partial _{y_{i}}u^{1}\text{ in }L^{2}(%
\widetilde{\Omega }\times Y),  \label{gradient:prop:limit}
\end{equation}%
together with the relation between $u^{1}$ and $(u^{0},\widetilde{u}^{1}),$%
\begin{equation}
\widetilde{u}^{1}=u^{1}+\sum_{j=1}^{n}y_{j}\partial _{x_{j}}u^{0}
\label{gradient:prop:tilde_u1}
\end{equation}%
as well as the fact that $u^{1}$ is $Y$-periodic. (iv) Once the
two-scale limit of $\partial _{x_{i}}u^{\varepsilon }$ has been
computed, it is used in the variational formulation
(\ref{model:diffusion:variational:eq}) in a manner that yields the
two-scale model (\ref{model:diffusion:two-scale:eq}). (v) Finally,
the homogenized model (\ref{model:diffusion:approx:eq}) is deduced
by expressing each microscopic derivatives $\partial _{y_{i}}u^{1}$
as a linear function of the macroscopic derivatives $(\partial
_{x_{j}}u^{0})_{j=1,...,n}$, the linear operator between them being
solution of a partial differential equation in the cell $Y$.

It would take too long to present in detail all of the above proof.
We chose to illustrate the transformation rules only on the third
and fourth steps because they require the implementation of most of
the useful concepts for the complete proof. In addition, we do not
show the periodicity of $u^{1}$
nor the relation (\ref{gradient:prop:tilde_u1}) between $u^{1}$ and $(u^{0},%
\widetilde{u}^{1}).$ Thus, the two examples discussed are to prove
the following propositions. The mathematical proofs are
reported in
Appendices \ref{appendix:gradient:math:proof} and \ref%
{appendix:diffusion:math:proof} respectively when their counterpart
formalized by rewriting rules and strategies are in Appendices \ref%
{appendix:gradient:maple:proof} and
\ref{appendix:diffusion:maple:proof}.

\begin{prop}
\label{gradient:approx:prop}For a sequence of functions
$u^{\varepsilon }:\Omega \rightarrow \mathbb{R}$ such that
$u^{\varepsilon }$ and $\partial
_{x_{i}}u^{\varepsilon }$ are bounded in the $L^{2}(\Omega )$-norm, if $%
Tu^{\varepsilon }$ has a formal expansion of the form%
\begin{equation}
Tu^{\varepsilon }=u^{0}+\varepsilon u^{1}+\varepsilon
\sum_{j=1}^{n}y_{j}\partial _{x_{j}}u^{0}+\varepsilon O(\varepsilon
), \label{approx:T:2:eq}
\end{equation}%
where $u^{0}$ is independent of $y$, the partial function $y\mapsto
u^{1}(.,y)$ is $Y$-periodic and $O(\varepsilon )$ tends to zero in $L^{2}(%
\widetilde{\Omega }\times Y)$ weak, then (\ref{gradient:prop:limit})
holds.
\end{prop}

\begin{prop}
\label{model:approx:prop}Assuming that the data of the stationary
heat equation (\ref{model:diffusion:variational:eq}) satisfy
\begin{equation}
Ta^{\varepsilon }=a^{0}\text{ and }Tf^{\varepsilon
}=f^{0}+O(\varepsilon ), \label{model:diffusion:assumption}
\end{equation}%
where $O(\varepsilon )$ tends to zero in $L^{2}(\widetilde{\Omega
}\times Y)$ and the solution $u^{\varepsilon }$ satisfies the
assumptions and the
conclusion of Proposition \ref{gradient:approx:prop}, then the pair $%
(u^{0},u^{1})$ is solution to the two-scale model (\ref%
{model:diffusion:two-scale:eq}).
\end{prop}

A number of mathematical tools are required to carry out the proofs.
Some of them are refered in the body of the paper so they are
described in the end of this section while the others are used only
in the detailed proofs and
thus are presented in the beginning of Appendix \ref%
{appendix:gradient:math:proof}.\\

In order to formalize the convergence concepts of sequences of
numbers and of functions, and to handle them within a computational
framework, we recall the usual set of computation rules of the
Landau notation $O(\varepsilon )$ and we also add redundant rules
for sums and integrals. As usual the computation rules on
the Landau notation apply from left to right only. 

\begin{eqnarray}
O(\varepsilon )+O(\varepsilon ) &=&O(\varepsilon ),\text{
}-O(\varepsilon )=O(\varepsilon ),\text{ }\sum O(\varepsilon
)=O(\varepsilon )\text{, }\int
O(\varepsilon )\text{ }dx=O(\varepsilon ),  \label{O1} \nonumber \\
O(\varepsilon )\ast O(\varepsilon ) &=&O(\varepsilon ),\text{
}\alpha \ast
O(\varepsilon )=O(\varepsilon )\text{ if }\alpha \text{ is independent of }%
\varepsilon ,\text{ and }\varepsilon =O(\varepsilon ).  \label{O2}
\end{eqnarray}%

This small system of ``\textit{axioms}''
defines what we call the \emph{convergence calculus }for the
rewriting rules. We observe that it does not include a mean for
distinguishing the various types of convergences, so for now that
distinction is left to the user. 
We shall also repeatedly use the additional property of
$O(\varepsilon )$ whereby
\begin{equation}
\int_{\Omega }g\text{ }O_{1}(\varepsilon )\text{ }dx=O(\varepsilon )
\label{O3}
\end{equation}%
as long as $g$ is a function uniformly bounded in $L^{2}(\Omega )$ and $%
\lim_{\varepsilon \rightarrow 0}||O_{1}(\varepsilon
)||_{L^{2}(\Omega )}=0.$
It results from the Cauchy-Scwharz inequality $\int_{\Omega }g$ $%
O_{1}(\varepsilon )$ $dx\leq ||g||_{L^{2}(\Omega
)}||O_{1}(\varepsilon )||_{L^{2}(\Omega )}\leq C$ $O(\varepsilon
)=O(\varepsilon ).$

The usual operations on variational formulation require to use the
extension of the rule of integration by parts to multidimensional
domains. The so-called \textit{Green formula} holds for sufficiently
regular functions $u$
and $v$ defined in a domain $\Theta \subset \mathbb{R}^{n},$%
\begin{equation}
\int_{\Theta }u\;\partial _{x_{i}}v\;dx=\int_{\partial {\Theta }%
}u\;v\,(n_{x})_{i}\;ds(x)-\int_{\Theta }v\partial _{x_{i}}u\;dx
\label{Green Formula}
\end{equation}%
for any $i=1,...,n$ where $n_{x}$ represents the outward pointing
unit normal of the hyper-surface volume element $ds(x)$ on the
boundary $\partial \Theta .$

\bigskip

Finally, functions such as $u$ and $v$ are usually considered as
elements of a vector space such as $L^{2}(\Omega )$. As such, it is
possible to define linear operators that apply to them. We recall
that an operator $L$ defined on a vector space is said to
be\textit{\ linear }if for any vectors $v,w$
and any scalar $\alpha $,%
\begin{align}
L(\alpha \;v)& =\alpha \;L(v)  \label{L1} \\
\text{ and }L(v+w)& =L(v)+L(w).  \label{L2}
\end{align}


%% file: rhomaple.tex
This section  defines a transformation language based on the three
notions of \emph{rule}, \emph{strategy} and \emph{transformation}. By a rule we
mean a classical rewrite rule. A strategy is a way to control how rules are
applied. Strategies can be combined to define strategies with a
finer control or a more powerful effect. We propose easy-to-remember names for
the most popular -- and indeed most useful -- strategy constructors and
combinators. The user can also extend the language with other strategies.

It is not obvious that any natural transformation of mathematical expressions
and models can be concisely expressed as a rewriting-based strategy, in a
natural way. Moreover what is exactly a strategy is not completely clear from
the literature, and the name of ``strategy'' for a formal transformation may
lead the user of our language to confusion. Therefore we consider an a priori
independent notion of \emph{transformation}. Transformations have the following
three features: (i) a transformation is reproducible, (ii) a transformation may
not progress, and (iii) a transformation may not terminate.
The reproducibility feature (i) means that each application of a given
transformation to a given expression produces the same effect: either it does
not terminate each time, or it produces each time the same expression. In
particular, this property excludes non-determinism from the notion of
transformation.
By contrast a strategy may be non-deterministic (see, e.g. \texttt{One[s]}
in~\cite{BBK_Tom07}).
It would be enriching to develop a complete theory of transformations but it
exceeds the scope of the present paper on rule-based transformations.

We propose an implementation of this transformation language as a new package
for the Maple$^{\mathsf{TM}}$ computer algebra system.
The package is named \ST, for ``symbolic transformations''.\footnote{The package
is available as an archive file upon request to the authors. It can be executed
on any machine running Maple$^{\mathsf{TM}}$.} We strongly rely on the
functional features of Maple$^{\mathsf{TM}}$ by providing rules, strategies and
transformations as Maple functions, possibly through higher-order functions
constructing them from another representation. Functions faithfully provide the
expected feature (i) of transformation reproducibility. Whether a transformation
output differs or not from its input (feature (ii))
is controlled by the Maple$^{\mathsf{TM}}$ mechanism of exceptions.
Feature (iii) is left under the responsability of users that can interrupt
execution with the Maple$^{\mathsf{TM}}$ function \texttt{timelimit}.

\subsection{Top rewriting}
The Maple$^{\mathsf{TM}}$ statement
\texttt{ruleName := [$l$,$r$]} declares the rewrite rule
$\rrule{l}{r}$ as a pair and assigns it the name \texttt{ruleName}.
The function \texttt{Transform} associates to any such pair the function
applying the corresponding rewrite rule at the top of any term:
Given a term $t$, the function application \texttt{Transform(ruleName)}$(t)$
applies the rewrite rule $\rrule{l}{r}$ to $t$, as defined in Definition
\ref{matching:def}. If the rule cannot
 be applied i.e. if \texttt{t} does not match its left side $l$,
then the exception  \texttt{"Fail"} is raised. This is the  standard
\emph{rewriting at the top} or \emph{top rewriting} strategy.

\begin{exmp}\label{top:exmp}
Consider the property
$\int v+w\;dx=\int v\;dx +\int w\;dx$
of linearity of the integral.
The rewrite rule corresponding to its application from left to right
can be defined with $\ST$ as the pair
\begin{verbatim}
 IntegralLinearity := [
  Integral(A_ + B_, C_),
  Integral(A, C) + Integral(B, C)];
\end{verbatim}
A convention in the package is that variable names end with "$\_$" in
order to distinguish them from constants.
In order to apply the  \texttt{IntegralLinearity} rule at the top to the term
\begin{verbatim}
 t := Integral(v(x)+w(x),x);
\end{verbatim}
we write \texttt{Transform(IntegralLinearity)(t)}. The resulting term is
\begin{verbatim}
 Integral(v(x),x) + Integral(w(x),x)
\end{verbatim}
\end{exmp}
 
\subsection{Elementary transformations}
The two elementary transformations \texttt{Identity} and \texttt{Fail} are
defined as follows.
\begin{displaymath}
 \begin{array}{|l l|}
\hline
  \texttt{Identity(t)}  &=_{\textit{def}}  \texttt{t} \\
  \texttt{Fail(t)}   &=_{\textit{def}}  \texttt{ error "Fail"}\\
\hline
\end{array}
\end{displaymath}
The first one has no effect, since it transforms \texttt{t} into itself. The
second one always fails and raises the exception \texttt{"Fail"}.
This exception is raised each time a transformation fails
transforming a term. What is a failure for a transformation has to
be defined for each transformation, as previously done for top
rewriting.

It is sometimes useful to consider the non-progress of a transformation as a
failure with the aim to handle this exception and enable other
transformations. This feature is realized by the transformation
combinator \texttt{IdentityAsFail} defined by
\begin{displaymath}
\begin{array}{|c l|}
\hline
  \texttt{IdentityAsFail(s)(t)}=_{\textit{def}}  & \texttt{ if  s(t) = t then
  Fail(t); else s(t);}\\
\hline
\end{array}
\end{displaymath}
for any transformation \texttt{s} and any term \texttt{t}.

Conversely, it is also convenient to hide  at some higher level the
exception \texttt{"Fail"} raised by a transformation \texttt{s}.
This is the purpose of the \texttt{FailAsIdentity} combinator:
\begin{displaymath}
\begin{array}{|c l|}
\hline
 \texttt{FailAsIdentity(s)(t)}=_{\textit{def}}  &\texttt{try s(t); catch "Fail" : t; }\\
\hline
\end{array}
\end{displaymath}


\subsection{Transformation combinators}
\label{combinators:sec}

In this section we define three \emph{transformation combinators}. They take
transformations as parameters and control their order of application. Thus they
help defining complex transformations by combination. They are generic in the
sense that they do not depend on the nature and structure of the terms they are
applied on.
They are defined as follows.
\begin{displaymath}
\begin{array}{|r l|}
\hline
 \texttt{LeftChoice([s${}_1$, $\ldots$, s${}_n$])(t)} &=_{\textit{def}}
 \texttt{try s$_1$(t);} \\ &  \hspace{0.8cm} \texttt{catch "Fail":}
     \texttt{LeftChoice([s${}_2$, $\ldots$,  s${}_n$])(t)}\\
\texttt{LeftChoice([\,])(t)} &=_{\textit{def}} \texttt{Fail(t);} \\
\hline
 \texttt{Comp([s${}_1$, $\ldots$,  s${}_n$])(t)} & =_{\textit{def}}
  \texttt{Comp([s${}_2$, $\ldots$,  s${}_n$])(s$_1$(t));} \\
\texttt{Comp([\,])(t)} &=_{\textit{def}} \texttt{t;} \\
\hline
\texttt{STNormalizer(s)(t)} &=_{\textit{def}} \texttt{if s(t) = t then t;} \\
         & \hspace{0.8cm} \texttt{else STNormalizer(s)(s(t));}\\
\hline
\end{array}
\end{displaymath}

The transformations \texttt{Left\-Choice([$s$])} and \texttt{Comp([$s$])} are
defined by induction on $n \geq 0$ for any sequence of transformations $s =
(\texttt{s$_i$})_{i= 1, \ldots, n}$. If $n = 0$, both do nothing. Otherwise, the
application of \texttt{Left\-Choice([$s$])} to the term \texttt{t} returns the
result \texttt{s$_i$(t)} of the application of the first transformation in the
sequence $s$ which succeeds on the term \texttt{t}. It reports a failure if no
one succeeds. When $n \geq 1$ the transformation \texttt{Comp([s$_1$, $\ldots$,
s$_n$])}  applies \texttt{s$_1$}, \texttt{s$_2$}, etc in sequence, until
application of \texttt{s$_n$} or a previous failure.
The transformation \texttt{STNormalizer(s)} iterates the application of the
transformation \texttt{s} until the latter fails or a fixed point is reached.
The transformation \texttt{STNormalizer(s)} fails if and only if the
transformation \texttt{s} fails during these iterated applications. To
avoid the failure of the transformation \texttt{s} when computing the normal
form of a term $t$ with respect to \texttt{s}, one should write
\texttt{STNormalizer(\texttt{FailAsIdentity}(s))($t$)}. The transformations
involving \texttt{STNormalizer} presents a risk of non termination and should
therefore be carefully employed. Notice that the above definition of
\texttt{STNormalizer} is just a specification. For more efficiency, the
implementation computes \texttt{s(t)} only once.

\ucomment{
\begin{exmp}
Let $t$ be the expression encoding the integral $\int v(x)+w(x)+z(x) \;dx$. The
statement

\texttt{STNormalizer(FailAsIdentity(Transform(IntegralLinearity)))($t$)}

\noindent applies the rule \texttt{IntegralLinearity} at the top of $t$
 until no further application is possible. The resulting term is
$\int v(x)\; dx + \int w(x)+z(x)\; dx$.
Notice that the statement

\texttt{STNormalizer(Transform(IntegralLinearity))($t$)}

\noindent produces the exception \texttt{"Fail"} because it applies
\texttt{IntegralLinearity} to $t$ once at the top, and then fails.
\end{exmp}
}



\subsection{Traversal transformations}
This section introduces transformations called \emph{traversal} or \emph{term}
transformations because they explore the structure of the
term they are applied on. We provide three traversal transformation
 constructors: \some, \STTop{} and \STBottom. By extension of a classical
 terminology in rewriting theory, a \emph{redex} of a term $t$ for a transformation $s$
 is a subterm of $t$ that can be transformed by $s$, i.e. on which the
 transformation $s$ does not fail.

The transformation \some(\texttt{s}) tries to apply the transformation
\texttt{s} to all the immediate subterms of a term \texttt{t}. It fails if all
these applications fail, or if \texttt{t} is a constant or a variable. It is
defined by
\begin{displaymath}
\begin{array}{|c l|}
\hline
\texttt{\some(s)(t)} =_{\textit{def}}
 & \texttt{if t = f(t$_1$,$\ldots$,t$_n$) then}\\
 & \hspace{0.3cm}
     \texttt{if } \forall i \in \{1,\ldots,n\}\;  \texttt{ s(t$_i$)
      fails, then}\\ & \hspace{0.6cm}
    \texttt{Fail(t)}; \\
 &\hspace{0.3cm} \texttt{else}\\
 & \hspace{0.6cm}
    \texttt{f(FailAsIdentity(s)(t$_1$),$\ldots$,FailAsIdentity(s)(t$_n$));}\\
 & \hspace{0.3cm} \texttt{fi;} \\
 & \texttt{else         // t is a constant or a variable} \\
 & \hspace{0.4cm} \texttt{Fail(t);} \\
 & \texttt{fi;} \\
 \hline
\end{array}
\end{displaymath}

\begin{exmp}
Consider again the rewrite rule \texttt{IntegralLinearity} of
Example \ref{top:exmp}, encoding the linearity of the integral, and
the term $t$ encoding $\int v(x)+w(x)\;dx$.
Notice that the statement

\texttt{Transform(IntegralLinearity)($t$+2)}

\noindent raises the exception \texttt{"Fail"} because the rewrite rule cannot
be applied at the top of the expression \texttt{$t$+2}. A solution is to
replace it with the statement

\some \texttt{(Transform(IntegralLinearity))($t$+2)}

\noindent which produces
$\int v(x)\;dx +\int w(x)\;dx + 2$ since the expression
\texttt{$t$+2} is viewed as the term $+(t,2)$.
\end{exmp}
The transformation \some{}  is  not very useful in
practice. Its main purpose is to shorten the definition of the other
traversal transformations,  namely \STTop{} and \STBottom.

The transformation \STTop(\texttt{s}) is very common in symbolic computation.
It applies the transformation $s$
once to all the redexes of $t$ for $s$ that are the closest ones to the root of
$t$, i.e. to the largest subterms of $t$ on which $s$ succeeds. In other words
the transformation \STTop{} traverses the term $t$ down from its root and tries to
apply $s$ to each traversed subterm. If the transformation $s$ succeeds on some subterm
$t'$ of $t$, then it is not applied to the proper subterms of $t'$. In
particular,  \STTop$(s)$ fails if and only if $s$ fails on all the subterms of $t$.
It can be formally defined by
\begin{displaymath}
\begin{array}{|c l|}
\hline
 \STTop\texttt{(s)}=_{\textit{def}} \texttt{LeftChoice}\texttt{([s,\some(\STTop(s))])} & \\
\hline
\end{array}
\end{displaymath}

%
%
%

\begin{exmp}
\label{OuterMost:ex}
Let
\begin{align*}
t=2+\underbrace{\int v(x) + 3(\underbrace{\int w(x)+g(x)\;dx}_{r_1})
\; dx}_{r_0}
\end{align*}
be a term with two redexes $r_0$ and $r_1$ for the rule of integral linearity.
Then

\texttt{\STTop(IntegralLinearity)($t$)}

\noindent gives the expression
\begin{align*}
2+\int v(x) \; dx + \int 3 (\int w(x)+g(x)\;dx) \; dx ,
\end{align*}
since the rule \texttt{IntegralLinearity} is only applied to the outermost redex
$r_0$ of $t$.
\end{exmp}

The strategy \STBottom$(s)$ works similarly, but in the
opposite direction, i.e. it traverses a term $t$ up from its smallest
subterms and tries to apply the strategy $s$ once to the smallest redexes
of $t$ for $s$. It is formally defined by
\begin{displaymath}
\begin{array}{|c l|}
\hline
 \STBottom\texttt{(s)}=_{\textit{def}} \texttt{LeftChoice}\texttt{([\some(\STBottom(s)),s])} & \\
\hline
\end{array}
\end{displaymath}

\begin{exmp}
\label{InnerMost:ex} For the term $t$ of  Example
\ref{OuterMost:ex}, the expression

\STBottom\texttt{(IntegralLinearity)}($t$)

\noindent applies the
rule \texttt{IntegralLinearity} only to the innermost redex $r_1$
of  $t$ and gives the expression
\begin{align*}
2+\int v(x) +  3(\int w(x)\;dx +\int g(x)\;dx) \; dx.
\end{align*}
\end{exmp}

%
%
%


We provide additional traversal transformations, namely \texttt{All},
\texttt{TopDown} and \texttt{BottomUp}. They are less useful
in practice, but we include them in \ST because they exist in other strategy
languages, e.g. the \textsc{Tom} strategy
language~\citep{BBK_Tom07}.

The transformation \texttt{All(s)} applies the transformation \texttt{s} to all
the immediate subterms of any term \texttt{t}, and it fails if and only if
 one of  the applications to the immediate subterms fails.
 It is defined by
\begin{displaymath}
\begin{array}{|c l|}
\hline
\texttt{All(s)(t)} =_{\textit{def}}
 & \texttt{if t = f(t$_1$,$\ldots$,t$_n$) then}\\
 & \hspace{0.6cm}
    \texttt{f(s(t$_1$),$\ldots$,s(t$_n$));}\\
 & \texttt{else         // t is a constant or a variable} \\
 & \hspace{0.6cm} \texttt{t;} \\
 & \texttt{fi;} \\
 \hline
\end{array}
\end{displaymath}

The main purpose of the transformation \texttt{All}
is to simplify the definition of the transformations \texttt{TopDown} and
\texttt{BottomUp}.

The transformation \texttt{TopDown(s)} tries to apply the transformation
\texttt{s} to all the subterms of any term \texttt{t}, at any depth, by starting with the
root of \texttt{t}. It fails when there is a  subterm of
\texttt{t} where \texttt{s} fails. It is defined by
\begin{displaymath}
\begin{array}{|c l|}
\hline
\texttt{TopDown(s)}=_{\textit{def}} \texttt{Comp}\texttt{([s,All(TopDown(s))])} & \\
\hline
\end{array}
\end{displaymath}
The transformation \texttt{BottomUp(s)} behaves similarly,
but works in the opposite direction, i.e. it
starts from the leaves (the smallest subterms) and goes up.
\begin{displaymath}
\begin{array}{|c l|}
\hline
\texttt{BottomUp(s)}=_{\textit{def}} \texttt{Comp}\texttt{([All(BottomUp(s)),s])} & \\
\hline
\end{array}
\end{displaymath}




%% file: advanced.tex
In this section we introduce some technical features of \ST, namely,
the ability to combine rewriting and procedural programming,
the matching modulo associativity and commutativity and the conditional
rewriting.

\subsection{Procedural programming and strict evaluation}
\label{procedural:program:subsec}
It is possible to combine rewrite rules and procedural programming in \ST. That is,
the right-hand side of the rewrite rules may contain calls of
Maple$^{\mathsf{TM}}$ predefined functions or of functions defined by the user. This feature is not
available  in pure rewriting languages, e.g. in Maude \citep{MaudeBook07},
but it is available in rewriting languages built upon a host language,
 e.g. $\rho$-log \citep{MP_RhoLog04} which is built on Mathematica
and Tom \citep{BBK_Tom07} which is built on Java.

When we declare   a rewrite rule whose right-hand side contains some function
calls a problem occurs: since Maple$^{\mathsf{TM}}$ is a strict evaluation language, it
completely evaluates all the sub-expressions of an expression before evaluating
the expression itself. Therefore it evaluates the function calls present in the
right-hand side of the rewrite rules at the \emph{declaration} of the rewrite
rule, whereas the expected behaviour is most often evaluation of function calls
 at the \emph{application} of the rewrite rule.
\begin{exmp}
  Let us consider the
following rule declaration:
\begin{verbatim}
 e := [L_,nops(L)];
\end{verbatim}
where \texttt{nops} is the Maple$^{\mathsf{TM}}$ function that computes the number of arguments
of \texttt{L}. When \texttt{L} is a list \texttt{nops(L)}
computes the length of \texttt{L}. The problem is that Maple$^{\mathsf{TM}}$
 immediately evaluates \texttt{nops(L)}
and the rule \texttt{e} becomes
\verb+[L_,1]+.
The expected behavior is that the function \texttt{nops} is
applied after the rule application, i.e. after instantiation of the variable
\texttt{L\_} by the substitution that arises from the matching
of \texttt{L\_} with a given term.
\end{exmp}

The  solution we propose is to write \texttt{DelayEval -> }$r$ instead of
$r$ for the right-hand side of a rewrite rule that contains function calls. In
the example, the declaration of \texttt{e} would be
\begin{verbatim}
 e := [L_, DelayEval -> nops(L)];
\end{verbatim}

The Maple$^{\mathsf{TM}}$ expression \texttt{x -> }$r$ denotes the
$\lambda$-term\footnote{We recall that a $\lambda$-term is either a variable
$x$, an abstraction $\lambda t.u$, or an application $(u \, v)$, where $u$ and
$v$ are two $\lambda$-terms.} $\lambda$\texttt{x}$.r$, and this solution works
for two reasons. First, the evaluation of $r$ in $\lambda$\texttt{x}$.r$ is
delayed until an argument is provided to this $\lambda$-term.
 Second, Maple$^{\mathsf{TM}}$
accepts the application of substitutions to $\lambda$-terms. With these two
ingredients we can correctly implement  the delayed evaluation in the rewrite
rules as follows. The implementation  of the application of the rule
 \texttt{[$l$, DelayEval -> $r$]} to the term $t$
is done via the following steps:
\begin{enumerate}
\item Compute a solution $\sigma$ to the matching problem $\match{l}{t}$,
\item Compute the application $\sigma(r)$ of $\sigma$ to $r$,
\item Return the application \texttt{(DelayEval -> $\sigma(r)$)(DelayEval)} of
the $\lambda$-term to the protected variable \texttt{DelayEval}.
\end{enumerate}
Since \texttt{DelayEval} is a protected variable in \ST and thus never appears
in $\sigma(r)$, the expression \texttt{(DelayEval -> $\sigma(r)$)(DelayEval)} is
$\beta$-reduced\footnote{The $\beta$-reduction is the reduction rule $(\lambda
t.u) v \leadsto_{_{\beta}} u[v/t]$, where $u[v/t]$ is the replacement of $t$
with $v$ in $u$.} by Maple$^{\mathsf{TM}}$ to $\sigma(r)$.

It is worth mentioning that enclosing with unevaluation
quotes \texttt{'...'} each function present in the right-hand
side $r$ of a rule, or using the parameter modifier \texttt{::uneval}, does
not give the required solution. If an unevaluated function
enclosed by the quotes appears as an argument of a term,
 then Maple$^{\mathsf{TM}}$ eliminates the quotes and  evaluates this
function. As a consequence, applying a substitution to the term $r$
provokes an early evaluation of the functions in $r$ enclosed by quotes.

\subsection{Extension to associative and commutative function symbols}
\ucomment{[[ Discussion :
toutes les solutions, ou seulement une. Si seulement une, déterministe ?
Reproductibilité. Revenir sur les exemples précédents contenant des +. Souvent
une seule solution suffit, est préférable. ]]}

When applying a rewrite rule $[l \rightarrow r]$ to a term $t$, and
 terms $l$ and/or $t$ contain associative and commutative function symbols,
such as $+$ or $*$, then the matching problem $\match{l}{t}$ and the rule application
have to be done \emph{modulo associativity and commutativity}.

The following  definition generalizes Definition \ref{matching:def}. It
defines pattern-matching and rule application modulo a theory as
  in  \citep{rhoCalIGLP-I+II-2001}.

\begin{defn}\label{matching_modulo:def}
Let $\mathbb{T}$ be an equational theory. For two terms $t$ and $t'$ in
$\mathcal{T(F,\mathcal{X})}$ the problem of finding substitutions $\sigma$
such that the equality  $\sigma(t')=t$ is a logical consequence of the axioms of
$\mathbb{T}$ is called a \emph{matching problem modulo} $\mathbb{T}$ and
is denoted $\matchth{t'}{t}{\mathbb{T}}$. 
Such a substitution $\sigma$ is called a \emph{solution} of the matching problem
$\matchth{t'}{t}{\mathbb{T}}$. The \emph{application} of the rewrite rule
$\rrule{l}{r}$ to a term $t$, denoted by $[\rrule{l}{r}](t)$, is the (possibly empty)
set $\{\sigma_1(r),\sigma_2(r),\ldots \}$
where each $\sigma_i$ is a solution of $\matchth{l}{t}{\mathbb{T}}$.
\end{defn}

If the theory $\mathbb{T}$ axiomatizes the associativity and
commutativity of some function symbols, then the set of solutions of the
matching problem is finite.  The associative and commutative symbols of \ST
presently are $*$, $+$, $\cup$ and $\cap$. The corresponding theory is denoted
by $AC$.

To deal with associative and commutative symbols the package \ST is equipped
with two matching functions. The function \texttt{Matching(t')(t)} returns only
one solution of the matching, always the same one, because transformations are
expected to be reproducible. The function \texttt{MatchingAll(t')(t)} returns
all the solutions. The first function is sufficient in many practical cases, but
the second one is sometimes useful. In particular it is necessary for the
application of conditional rules (see next section).

The present implementation of these two functions in \ST does not include
 optimizations suggested in the literature.
In \citep{WRS11}, the authors suggested a more flexible
approach. They provided a \emph{lazy} matching algorithm modulo associativity and
commutativity. \emph{Lazy} means that the algorithm produces only a first
solution and a way to get the other ones. We plan to integrate this lazy
algorithm in a future version of \ST.

\subsection{Conditional rewriting}
In \ST rewrite rules can be conditional. A conditional rule is a rewrite rule
$\rrule{l}{r}$ and a condition $c$ on the variables in $l$.
In \ST a conditional rule is
declared as a list \verb++[$l$\verb+,+$r$\verb+,+$c$\verb+]+  of three elements,
where the third element is the condition. Notice that the \texttt{DelayEval}
mechanism is usually required in the condition when it contains function calls.

\begin{exmp}
The linearity property (\ref{L1}) of an operator $B$ can be encoded by
the following conditional rewrite rule replacing $B(\alpha x)$ by $\alpha B(x)$
if $\alpha$ is a scalar:
\begin{verbatim}
 [B(Alpha_*X_), Alpha*B(X), DelayEval -> IsScalar(Alpha)].
\end{verbatim}
\end{exmp}

The semantics of the application
\texttt{Transform([}$l$\texttt{,}$r$\texttt{,}$c$\texttt{])(}$t$\texttt{)}  of
the conditional rule
\verb++[$l$\verb+,+$r$\verb+,+$c$\verb+]+ to a term $t$, where $c$ is a Boolean
condition, is a term $\sigma(r)$ where $\sigma$ is a solution of the  matching
problem $\matchth{l}{t}{AC}$ such that $\sigma(c)$ holds, or raises the exception
\texttt{"Fail"} if there is no such solution.



%% file: gradient-input-explanations.tex
As typical examples, we consider the mathematical proofs reproduced in
Appendices~\ref{appendix:gradient:math:proof}
and~\ref{appendix:diffusion:math:proof}. They are respectively formalized in
Appendices~\ref{appendix:gradient:maple:proof} and
\ref{appendix:diffusion:maple:proof} as sequences of transformations with the
$\ST$ language. Appendix~\ref{appendix:rules} shows the collection of rules
and transformations used in the proofs. The
present section explains this formalization.


%% file: encoding.tex
\subsection{Mathematical operators}
The algebraic properties of the integration, derivation, and general summation
operators are encoded with \ST in Appendix \ref{appendix:rules}, Section
\ref{IntegralRules:sec}. Two versions of the Green rule are given
in Appendix \ref{appendix:rules}, Section \ref{GreenRule:maple}: the usual rule
\texttt{GreenRule}, and a parametrized conditional one \texttt{CondGreenRule}.
The pattern \texttt{patt} is used to ensure that the Green rule is applied in
the right way.

\subsection{About linearity}
We recall that an operator $L$ is said to be linear if
(\ref{L1}) and (\ref{L2}) hold. The linear operators used in the proofs are
$T$, $T^{*}$, $B$, $\int_{\Omega}$, $\partial_{x}$ and  $\sum_{i}$. At the
present level of formalization, scalars are not distinguishable from other symbolic expressions.
As a consequence Eq.~(\ref{L1})  cannot be turned into a general rewrite rule:
otherwise this rule could also produce the unexpected term $v\; L(\alpha)$ from
$L(\alpha \; v)$. We presently address this issue by writing one rewrite rule
for each term $\alpha$ of interest in the proofs under consideration. It only concerns Steps
2 and 5 in Appendix~\ref{appendix:diffusion:math:proof}.

On the contrary  Eq.~(\ref{L2}) can be safely expressed by a rewrite
rule. However the two-scale transform manipulates many linear
operators and it is tedious  to define a rewrite rule that expresses
the linearity property (\ref{L2}) for each operator. Therefore  we
provide the  generic constructor \texttt{Linearity(n,fun,t)}, where
\texttt{fun} is a function and \texttt{t} is a term with \texttt{n}
(underscored) variables or more. This constructor generates a
rewrite rule that states that the operator \texttt{t} is linear with
respect to its $\texttt{n}^{\textrm{th}}$ variable, in the sense of
\texttt{fun}. Notice that \texttt{fun} is usually the $+$ function
provided by Maple$^{\mathsf{TM}}$. However it is possible to have
different $+$ functions for different vector spaces. For instance,
the Maple$^{\mathsf{TM}}$ expressions
\begin{center}
\verb!Linearity(2,x->y->x+y,Integral(Omega_,_,Z_));!
\end{center}
and
\begin{center}
\verb!Linearity(1,x->y->x+y,T(_));!
\end{center}
respectively express the linearity
property (\ref{L2}) of the \texttt{Integral} and \texttt{T} operators with respect to $+$.
Their evaluation respectively produces the rule
\begin{center}
\verb![Integral(Omega_,X_+Y_,Z_), Integral(Omega,X,Z)+Integral(Omega,Y,Z)]!
\end{center}
and
\begin{center}
\verb![T(X_+Y_), T(X)+T(Y)]!.
\end{center}

For a given operator, the two rewrite rules that correspond to
 (\ref{L1}) and (\ref{L2}) can be merged together using transformation
 combinators.
The collection of linear operators as well as their linearity properties
 is stored in an array named \texttt{LinearityOf} and defined in the file
 \texttt{integral.mpl} reproduced in Appendix \ref{appendix:rules}, Section
\ref{IntegralRules:sec}.

\subsection{Convergence calculus}
\label{convergence:sec}
\input convergence.tex

\subsection{Two-scale calculus}
The algebraic properties of the two-scale operators have been stated and
mathematically proved in Appendix A of~\citep{LenSmi07}. They are formulated as
rewriting rules in Appendix~\ref{appendix:rules}, Section
\ref{two-scale:mpl:annex}.

The two-scale limit of the gradient operator corresponds to Eq.~(74) of
\citep{LenSmi07}. This equality is proved with \ST  in
Appendix~\ref{appendix:gradient:maple:proof}. The mechanism of the formal proof
is to reduce the left- and the right-hand sides of the
 equality~\ref{gradient:approxim:eq}, and then to show that their reduced forms
 are equal up to $O(\varepsilon)$.

The  formal  derivation of the two-scale  model of the stationary heat equation
 with \ST is reproduced in Appendix
\ref{appendix:diffusion:maple:proof}. There, the transformation steps are
applied to the term \texttt{LHST} $-$ \texttt{RHST}, where \texttt{LHST} (resp.
\texttt{RHST}) is the left- (resp. right-) hand side term of the heat equation
(\ref{model:diffusion:variational:eq}). The two-scale model derivation uses the
weak convergence property of the derivative operator (Step 6 in Appendix
\ref{appendix:diffusion:math:proof}) as a lemma. In the formal proof this lemma
is written ``by hand'' as a rewriting rule after being formally
proved.


\subsection{Final remarks}
Some proof steps require to apply some equation $A=B$ from left to right
and from right to left. However, including  the two rewriting rules $A \rightarrow
B$ and $B \rightarrow A$ in the same strategy may induce non-termination.
This is typically the case for the linearity properties.
In this case, non-termination is avoided by the introduction of a more
specialized version of the rules corresponding to the second
orientation, such as:
\begin{verbatim}
AdhocSimplify := [Integral(Omega_,C_*SUM(F_,J_,D_),X_),
                  SUM(Integral(Omega,C*F,X),J,D)];
\end{verbatim}


Notice that the proofs use two Maple$^{\mathsf{TM}}$ functions:
\texttt{has(expr1,expr2)} that returns true if \texttt{expr2} is a subexpression
of \texttt{expr1}, and \texttt{Evala(Expand(expr))} that
  expands products and powers of rational functions with algebraic coefficients.

We have also developed a formal proof of the property (\ref{gradient:cond:u0})
that the two-scale weak limit $u^{0}$ of $Tu^{\varepsilon}$ is independent of
$y$. This proof is not reproduced here. Instead, (\ref{gradient:cond:u0}) is
encoded as a lemma (See Appendix~\ref{lemmas:maple:appendix}) used in the
main proofs. The hypotheses of these proofs are gathered in the file
\texttt{hypothesis.mpl} reproduced in
 Appendix \ref{appendix:rules}, Section \ref{hypotheses:maple:appendix}.



%% file: convergence.tex
The notion of convergence has been introduced in Section~\ref{prelim:sec}. 
The theory of convergence (\ref{O2}) and (\ref{O3}) corresponds to the following
rewrite rules:
\begin{eqnarray}
&&\rrule{O(\varepsilon)+O(\varepsilon)}{O(\varepsilon)}, \label{sum:rule} \\
&&\rrule{\sum_{i=1}^{n}O(\varepsilon)}{O(\varepsilon)}, \label{big:sum:rule} \\
&&\rrule{\int_{\Omega}O(\varepsilon)\;dx}{O(\varepsilon)},\text{ and }
            \label{integral:rule} \\
&&\rrule{z * O(\varepsilon)}{O(\varepsilon)} \label{const:rule} 
\end{eqnarray}
for a term $z$ bounded with respect to $\varepsilon$. These rules are combined
in a strategy defined by
\begin{verbatim}
ConvergenceStrategy :=
  STNormalizer(
    LeftChoice([
      Outermost(OEpsilonSum),
      Outermost(OEpsilonSUM),
      Outermost(OEpsilonIntegral),
      Outermost(OEpsilonConst)
    ])
  );
\end{verbatim}
where \texttt{OEpsilonSum}, \texttt{OEpsilonSUM}, 
\texttt{OEpsilon\-Integral} and \texttt{OEpsilonConst} respectively encode the rewriting rules
(\ref{sum:rule}), (\ref{big:sum:rule}), (\ref{integral:rule}) and (\ref{const:rule}).
They are defined in the file \texttt{conver\-gence.mpl} reproduced in
Appendix \ref{appendix:rules}, Section~\ref{convergence:rules:sec}. The result
\texttt{ConvergenceStrategy} is a powerful strategy that reduces $O(\varepsilon)$ terms as much as possible. In the present
case it can be shown that it always terminates and thus can be systematically
applied after each tranformation step. 

We now present and address
a problem arising when embedding this strategy within a computer
algebra system with strict evaluation.
If the notion of a function that tends to $0$ when $\varepsilon$ tends to $0$ is
represented by the Maple$^{\mathsf{TM}}$ expression $\OEps$, then Maple$^{\mathsf{TM}}$ simplifies any
expression $\OEps - \OEps$ to zero, whereas $\OEps - \OEps$ should be
simplified to $\OEps$.
The solution we suggest consists in considering the term $\OEpsi{i}$ instead of
$\OEps$, where $i$ is a \emph{fresh} index. The term ``fresh'' means that the same index
has never been produced before to construct such a term. This solution
 is natural because it basically relies  on the mathematical semantics of the
 term $\OEps$.  That is, two occurrences of $\OEps$ are two different
functions, and it is natural to distinguish them using two different indexes.
Technically, we provide a function \texttt{FreshIndex()} that returns a new index
at each call. Moreover each occurrence of $\OEps$ in the right-hand side of a
rewrite rule is replaced by $\OEpsi{\texttt{Fresh\-Index()}}$.

%% file: related.tex
The  theoretical basis for this work   is the
deterministic fragment of the $\rho_{_{AC}}$-calculus ~\citep{rhoCalIGLP-I+II-2001}, 
where $AC$ is the theory axiomatizing the associativity and commutativity
of the symbols $+,*, \cup$ and  $\cap$. 
 The $\rho_{_{\mathbb{T}}}$-calculus, where $\mathbb{T}$ is an equational theory,
  is an extension of the $\lambda$-calculus;
one abstracts on a pattern rather than on  a single variable. 
The abstraction
mechanism is based on the rewrite rule $\rrule{l}{r}$, also  viewed as a
$\rho$-term. Notice that when $l$ is just a variable $x$, this $\rho$-term
corresponds to the $\lambda$-term $\lambda x.\;r$. 
Moreover, the $\rho_{_{\mathbb{T}}}$-calculus considers higher-order terms, i.e.
terms that may contain abstractions and  rule applications.

When an abstraction $\rrule{l}{r}$ is applied to a $\rho$-term $t$, which is
denoted by $[\rrule{l}{r}]t$, the matching mechanism is based on the binding
 of the free variables of $l$ to the appropriate subterms of $t$. 
This matching is  done modulo the  theory
$\mathbb{T}$. The latter is often expressed by algebraic axioms such as
associativity and/or commutativity.
 In  \ST  both the left-hand side term of a rule  and the term under
rule application  corresponds to first-order $\rho$-terms, 
i.e. they  contain neither abstractions nor rule applications. 
However the right-hand side of a rule may be a higher-order  $\rho$-term,
since it may contain   function calls. Those functions 
are nothing but $\lambda$-terms. Handling the priority between rule application
and $\beta$-reduction is explained in the steps (1), (2) and (3) in 
Section \ref{procedural:program:subsec}. Despite 
the fact that strategies can be encoded with the $\rho_{_{\mathbb{T}}}$-calculus 
\citep{RewriteStrat_CHK2003} by means of  some constructors, we preferred
encoding the strategies in \ST by means of Maple functions for the sake 
of efficiency.
Finally, we notice that in the \ST language, if  a rule cannot be applied
 to a term then  the exception \texttt{Fail} is raised. This
    makes a subtle difference with the semantics of the 
$\rho_{_{\mathbb{T}}}$-calculus
 that consists in returning an empty set in this case. The problem  of the
 $\rho_{_{\mathbb{T}}}$-calculus  approach is that we can not distinguish between
an empty set which is a mathematical term that could arise  from the
 symbolic transformations, and the empty set which denotes the failure of the
 application of a rule. 

The proposed transformation language does not claim for originality. It is
deliberately an adaptation for Maple of popular strategy languages such as
$\rho$-log~\citep{MP_RhoLog04} or \textsc{Tom}~\citep{BBK_Tom07}. But,
departing from \textsc{Tom} which extends an host language with an additive
syntax, our transformation language smoothly integrates with standard Maple
functions. Consequently, the Maple programmer learns it quickly, and 
 is free to mix
function- and rule-based programming styles. Moreover all the features of
her development environment (such as refactoring, code completion, dependency
analyses, etc) are preserved for free.

The closest implementation is $\rho$-log, a package developed upon the 
 advanced rewriting kernel of Mathematica.
It supports non-deterministic and  conditional rewriting.  
  The main drawback of   $\rho$-log is that it 
 considers the non-applicability of a rule as the identity. Technically
 speaking, the  strategy \texttt{FailAsIdentity} is implicitly applied 
 to all the transformations. However, when a transformation returns 
 the same term given as an input, we do not know if this transformation
 fails or it performs some modifications and then returns the same term. 
Moreover in $\rho$-log it is not possible, at least in a straightforward way,
to do higher-order rewriting, since the rewriting rules are not directly
accessible to the user: They are declared by means of the constructor 
\texttt{DeclareRule}.


%% file: conclusion.tex
Our main motivation for the development of a transformation language in Maple$^{\mathsf{TM}}$
was to facilitate the design of
 the  MEMSALab software  
dedicated to the automatic derivation of multi-scale models. However \ST is a general tool that can be 
 used by  Maple$^{\mathsf{TM}}$ programmers and  mathematicians
 in  the formalization of equational reasoning.  It makes it possible 
to express  rule-based symbolic computations in a concise and natural way,
thus providing a good  guarantee of the correctness of the formal proofs
  with respect to their hand-written counterparts. Since the \ST package is
written in Maple$^{\mathsf{TM}}$, it obviously does not extend the expressivity of the
Maple$^{\mathsf{TM}}$ language, but it clearly increases readability and
conciseness. Although this paper presents an implementation in Maple$^{\mathsf{TM}}$$^{\mathsf{TM}}$, the transformations
presented here could easily be developed in a similar way in any functional
language.

The transformation language \ST allows the derivation of  the weak two-scale limit of 
the derivative operator and the two-scale model of the stationary heat equation
  at the same ``level'' as the   mathematical derivations. 
 The word ``level'' covers three aspects:  the formal and the hand-written 
 proofs  have almost the same size,  they follow the same  steps, and  the
 strategy term written at each step of the formal proofs is a natural
 formalization of its mathematical counterpart.
The \ST package is used in 
\citep{EuroSim11,CFM11} to formally derive the two-scale model of the stationary 
heat equation in a region composed of a thin part and a part with periodically distributed holes.

For a more scalable treatment of linearity we plan in a near future to detect
the scalar nature of terms by assigning a type to each expression. More
generally a type system for  mathematical expressions
 is under way. We plan to transform each proof into a module whose
 execution produces a parametrized rewrite rule. The latter can be
instantiated and applied in other proofs.


%% file: GradientProof/gradient.tex
This section is devoted to the proof of Proposition \ref{gradient:approx:prop}. 
We start with some reminders of mathematics
that complement those in Section \ref{example:sec}.

Here the two-scale transform $T$ can be viewed as a linear
continuous operator from $L^{2}(\Omega )$ into
$L^{2}(\widetilde{\Omega }\times Y)$, as
such its adjoint $T^{\ast }$ is a linear continuous operator from $L^{2}(%
\widetilde{\Omega }\times Y)$ into $L^{2}(\Omega )$ defined by
\begin{equation}
\int_{\widetilde{\Omega }\times Y}T(u)\text{ }v\text{ }dxdy=\int_{\Omega }u%
\text{ }T^{\ast }(v)\text{ }dx\text{ for any }u\in L^{2}(\Omega )\text{ and }%
v\in L^{2}(\widetilde{\Omega }\times Y).  \label{T:adjoint:eq}
\end{equation}%
The two-scale transform can also be defined on integrable functions
and it satisfies the property%
\begin{equation}
T(u\text{ }v)=T(u)T(v)\text{ for any }u,v\in L^{2}(\Omega ).
\label{T:property}
\end{equation}%
Then, we define the so-called \textit{regularized inverse two-scale transform%
} $B:L^{2}(\widetilde{\Omega }\times Y)\rightarrow L^{2}(\Omega )$
by
\begin{equation*}
B(v)(x)=v(x,x/\varepsilon ).
\end{equation*}%
It can be easily checked that $B$ is a linear operator. The partial
derivatives of $B(v)$ for any sufficiently regular function $v$ can be
derived by applying the chain rule%
\begin{equation}
\partial _{x_{i}}B(v)=B(\partial _{x_{i}}v)+\frac{1}{\varepsilon }B(\partial
_{y_{i}}v).  \label{B:derivative}
\end{equation}%
It is also useful to know how the null condition of a function $v(x,y)$ on the
boundary $\partial (\widetilde{\Omega }\times Y)$  is transfered
to its range by $B$:%
\begin{equation}
\text{If }v=0\text{ on }\partial (\widetilde{\Omega }\times Y)\text{ then }%
B(v)=0\text{ on }\partial \Omega .  \label{B:property}
\end{equation}
In the following lemma (admitted), $O(\varepsilon )$ denotes any
function that vanishes in the $L^{2}(\Omega )$-norm when
$\varepsilon $ tends to zero.

\begin{lem}
\label{Approx T*}
The operator $B$ is a zero-order
approximation of the adjoint operator $T^{\ast }$ in the
sense
that %
\begin{equation}
T^{\ast }(v)-B(v)=O(\varepsilon )  \label{B:approximation:1}
\end{equation}
for any sufficiently regular and $Y$-periodic function $v$. 
 Moreover, $B(v)$ can be approximated at the
first-order by
\begin{equation}
B(v)=T^{\ast }(v+\varepsilon \sum_{j=1}^{n}y_{j}\partial
_{x_{j}}v)+\varepsilon O(\varepsilon ).  \label{B:approximation:2}
\end{equation}
\end{lem}

\bigskip

In the following, for simplicity we write $u$ instead of
$u^{\varepsilon }$. We shall prove Proposition
\ref{gradient:approx:prop} or equivalently, by the density of the
set $\mathcal{C}_{0}^{\infty }(\widetilde{\Omega }\times Y)^{n}$ of
infinitely continuously differentiable functions will all
derivatives vanishing on $\partial (\widetilde{\Omega }\times Y)$ in
the set
$L^{2}(\widetilde{\Omega }\times Y)^{n}$, that%
\begin{equation}
\underbrace{\sum_{i=1}^{n}\int_{\widetilde{\Omega }\times
Y}T(\partial _{x_{i}}u)v_{i}\;dxdy}_{\Psi
}=\sum_{i=1}^{n}\int_{\widetilde{\Omega }\times Y}(\partial
_{x_{i}}u^{0}+\partial _{y_{i}}u^{1})v_{i}\;dxdy+O(\varepsilon )
\label{gradient:approxim:eq}
\end{equation}
for any $v=(v_{1},...,v_{n})\in \mathcal{C}_{0}^{\infty }(\widetilde{\Omega }%
\times Y)^{n}.$

\begin{itemize}
\item \textbf{Step 1.} Applying the definition of $T^{\ast }$ to the left-hand
side $\Psi$ of (\ref{gradient:approxim:eq}) yields
\begin{equation*}
\Psi =\sum_{i=1}^{n}\int_{\Omega }\partial _{x_{i}}u\,T^{\ast
}(v_{i})\text{ }dx.
\end{equation*}

\item \textbf{Step 2.} From the approximation (\ref{B:approximation:1}) of $%
T^{\ast }(v_{i})$ by $B(v_{i})$, the linearity of integral, the
boundedness of
$||\partial _{x_{i}}u||_{L^{2}(\Omega )}$ and the property (\ref{O3}) of $%
O(\varepsilon )$ we get

\begin{align*}
\Psi & =\sum_{i=1}^{n}\int_{\Omega }\partial _{x_{i}}u\,B(v_{i})\text{ }%
dx+\sum_{i=1}^{n}\int_{\Omega }\partial _{x_{i}}u\,O(\varepsilon
)\text{ }dx
\\
& =\sum_{i=1}^{n}\int_{\Omega }\partial _{x_{i}}u\,B(v_{i})\text{ }%
dx+O(\varepsilon ).
\end{align*}

\item \textbf{Step 3.} Then, we apply the Green formula (\ref{Green Formula}%
) and get
\begin{equation*}
\Psi =\sum_{i=1}^{n}\int_{\partial \Omega }u\text{
}B(v_{i})(n_{x})_{i}\text{
}ds(x)-\sum_{i=1}^{n}\int_{\Omega }u\;\partial _{x_{i}}B(v_{i})\text{ }%
dx+O(\varepsilon ),
\end{equation*}%
and the terms on the boundary are removed thanks to Property (\ref%
{B:property}),%
\begin{equation*}
\Psi =-\sum_{i=1}^{n}\int_{\Omega }u\;\partial _{x_{i}}B(v_{i})\text{ }%
dx+O(\varepsilon ).
\end{equation*}

\item \textbf{Step 4.} From the expression (\ref{B:derivative}) applied to
the partial derivatives of $B(v_{i})$ and by linearity of integral,
\begin{equation*}
\Psi =-\sum_{i=1}^{n}\int_{\Omega }u\;\big(B(\partial _{x_{i}}v_{i})+\frac{1%
}{\varepsilon }B(\partial _{y_{i}}v_{i})\big)\text{
}dx+O(\varepsilon ),
\end{equation*}%
and
\begin{equation*}
\Psi =-\sum_{i=1}^{n}\big[\underbrace{\int_{\Omega }uB(\partial
_{x_{i}}v_{i})\text{ }dx}_{\Psi _{1}}+\underbrace{\int_{\Omega }\frac{1}{%
\varepsilon }uB(\partial _{y_{i}}v_{i})\text{ }dx}_{\Psi _{2}}\big]%
+O(\varepsilon ).
\end{equation*}

\item \textbf{Step 5.} We apply (\ref{B:approximation:1}) and (\ref%
{B:approximation:2}) to approximate $B(\partial _{x_{i}}v_{i})$ at
the zero-order and $B(\partial _{y_{i}}v_{i})$ at the first-order
together with
the rule (\ref{O3}) and thus get%
\begin{equation*}
\Psi _{1}=\int_{\Omega }u\;T^{\ast }(\partial
_{x_{i}}v_{i})\,dx+O(\varepsilon )
\end{equation*}%
and
\begin{equation*}
\Psi _{2}=\int_{\Omega }\frac{1}{\varepsilon }u\;\big[T^{\ast
}(\partial _{y_{i}}v_{i}+\varepsilon \sum_{j=1}^{n}y_{j}\partial
_{x_{j}}\partial _{y_{i}}v_{i})\big]dx+O(\varepsilon ).
\end{equation*}%
Thanks to the linearity of $T^{\ast }$,
\begin{equation*}
\Psi _{2}=\int_{\Omega }\;u\;T^{\ast }(\frac{1}{\varepsilon
}\partial
_{y_{i}}v_{i}+\sum_{j=1}^{n}y_{j}\partial _{x_{j}}\partial _{y_{i}}v_{i})%
\text{ }dx+O(\varepsilon ).
\end{equation*}%
Grouping $\Psi _{1}$ and $\Psi _{2},$%
\begin{equation*}
\Psi =-\sum_{i=1}^{n}\big[\int_{\Omega }u\;T^{\ast }(\partial
_{x_{i}}v_{i})\,dx+\int_{\Omega }\;u\;T^{\ast }(\frac{1}{\varepsilon }%
\partial _{y_{i}}v_{i}+\sum_{j=1}^{n}y_{j}\partial _{x_{j}}\partial
_{y_{i}}v_{i})\text{ }dx\big]+O(\varepsilon ).
\end{equation*}

\item \textbf{Step 6.} From the definition of the dual operator $T^{\ast }$
of $T$,

\begin{equation*}
\Psi =-\sum_{i=1}^{n}\big[\int_{\widetilde{\Omega }\times
Y}T(u)\;(\partial
_{x_{i}}v_{i})\,dxdy+\int_{\widetilde{\Omega }\times Y}\;T(u)\;(\frac{1}{%
\varepsilon }\partial _{y_{i}}v_{i}+\sum_{j=1}^{n}y_{j}\partial
_{x_{j}}\partial _{y_{i}}v_{i})\text{ }dxdy\big]+O(\varepsilon ).
\end{equation*}%
After expanding and applying linearity of integral,
\begin{eqnarray*}
\Psi 
&=  -\sum_{i=1}^{n}\big[ &
       \int_{\widetilde{\Omega }\times Y}T(u)\;(\partial_{x_{i}}v_{i})\,dxdy
       +\int_{\widetilde{\Omega }\times Y}\;
         T(u)\frac{1}{\varepsilon}\partial_{y_{i}}v_{i}\,dxdy \\
&  & +\int_{\widetilde{\Omega }\times
Y}T(u)\sum_{j=1}^{n}y_{j}\partial _{x_{j}}\partial _{y_{i}}v_{i}\text{ }dxdy%
\big]+O(\varepsilon).
\end{eqnarray*}

\item \textbf{Step 7. } We use the zero-order approximation%
\begin{equation}
Tu^{\varepsilon }=u^{0}+\varepsilon O(\varepsilon ),
\label{approx:T:1:eq}
\end{equation}
and the first-order approximation, both deduced from
(\ref{approx:T:2:eq}), respectively in the first and third integrals
and in the second integral to get
\begin{align*}
\Psi =& -\sum_{i=1}^{n}\big[\int_{\widetilde{\Omega }\times
Y}(u^{0}+O(\varepsilon ))\;\partial _{x_{i}}v_{i}\;dxdy \\
& +\int_{\widetilde{\Omega }\times Y}\big(u^{0}+\varepsilon
u^{1}+\varepsilon \sum_{j=1}^{n}y_{j}\partial
_{x_{j}}u^{0}+\varepsilon
O(\varepsilon )\big)\frac{1}{\varepsilon }\partial _{y_{i}}v_{i}\;dxdy \\
& +\int_{\widetilde{\Omega }\times Y}(u^{0}+O(\varepsilon
))\;\sum_{j=1}^{n}y_{j}\partial _{x_{j}}\partial _{y_{i}}v_{i}\;dxdy\big]%
+O(\varepsilon ).
\end{align*}%
After simplification
\begin{align*}
\Psi =& -\sum_{i=1}^{n}\big[\int_{\widetilde{\Omega }\times
Y}u^{0}\;\partial _{x_{i}}v_{i}\;dxdy+\int_{\widetilde{\Omega }\times Y}%
\frac{1}{\varepsilon }u^{0}\partial _{y_{i}}v_{i}\;dxdy \\
& +\int_{\widetilde{\Omega }\times Y}u^{1}\partial
_{y_{i}}v_{i}\;dxdy+\sum_{j=1}^{n}\int_{\widetilde{\Omega }\times
Y}y_{j}\partial _{x_{j}}u^{0}\,\partial _{y_{i}}v_{i}\;dxdy \\
& +\int_{\widetilde{\Omega }\times
Y}u^{0}\sum_{j=1}^{n}y_{j}\partial _{x_{j}}\partial
_{y_{i}}v_{i}\;dxdy\big]+O(\varepsilon ).
\end{align*}

\item \textbf{Step 8. } We apply the following instance of the Green formula
to the second subterm,%
\begin{equation*}
\int_{Y}u^{0}\partial _{y_{i}}v_{i}\;dy=\int_{\partial
Y}u^{0}v_{i}n_{y_{i}}\;ds(y)-\int_{Y}\partial
_{y_{i}}u^{0}\,v_{i}\;dy,
\end{equation*}%
where $n_{y}$ stands for the unit outward normal vector to the boundary $%
\partial Y$ of $Y$. Remarking that $\partial _{y_{i}}u^{0}$ vanishes%
\footnote{%
In the formal proof this simplification is done after the successive
applications of Green rule and the elimination of the boundary
terms, i.e. at the end of step 9.} and that $v$ vanishes on
$\partial Y$, the second subterm vanishes and we obtain

\begin{align*}
\Psi =& -\sum_{i=1}^{n}\big[\int_{\widetilde{\Omega }\times
Y}u^{0}\;\partial _{x_{i}}v_{i}\;dxdy+\int_{\widetilde{\Omega
}\times
Y}u^{1}\partial _{y_{i}}v_{i}\;dxdy \\
& +\int_{\widetilde{\Omega }\times Y}\sum_{j=1}^{n}y_{j}\partial
_{x_{j}}u^{0}\,\partial _{y_{i}}v_{i}\;dxdy+\int_{\widetilde{\Omega
}\times
Y}u^{0}\sum_{j=1}^{n}y_{j}\,\partial _{x_{j}}\partial _{y_{i}}v_{i}\;dxdy%
\big]+O(\varepsilon ).
\end{align*}

\item \textbf{Step 9. }Similarly, repeating the Green formula application
until no derivative is left on the test function $v,$%
\begin{eqnarray*}
\Psi & =-\sum_{i=1}^{n} \Big [ &\int_{\partial \widetilde{\Omega
}\times Y}u^{0}\text{
}v_{i}(n_{x})_{i}\;ds(x)dy-\int_{\widetilde{\Omega }\times
Y}\partial _{x_{i}}u^{0}\,v_{i}\;dxdy \\
&&+\int_{\widetilde{\Omega }\times \partial
Y}u^{1}v_{i}n_{y_{i}}\;dxds(y)-\int_{\widetilde{\Omega }\times
Y}\partial
_{y_{i}}u^{1}\,v_{i}\;dxdy \\
&&+\sum_{j=1}^{n}\int_{\widetilde{\Omega }\times \partial
Y}y_{j}\,\partial
_{x_{j}}u^{0}\,v_{i}n_{y_{i}}\;dxds(y)-\sum_{j=1}^{n}\int_{\widetilde{\Omega
}\times Y}\partial _{y_{i}}(y_{j}\partial _{x_{j}}u^{0})\,v_{i}\;dxdy \\
&&+\sum_{j=1}^{n}\int_{\partial \widetilde{\Omega }\times \partial
Y}u^{0}v_{i}n_{y_{i}}\,y_{j}n_{x_{j}}\;ds(x)ds(y) \\
&& -\sum_{j=1}^{n}\int_{%
\partial \widetilde{\Omega }\times Y}\partial
_{y_{i}}(y_{j}u^{0})\,n_{x_{j}}v_{i}\text{ }ds(x)dy \\
&&-\sum_{j=1}^{n}\int_{\widetilde{\Omega }\times \partial
Y}y_{j}\partial
_{x_{j}}u^{0}\text{ }v_{i}n_{y_{i}}\text{ }dxds(y)\\
&& +\sum_{j=1}^{n}\int_{%
\widetilde{\Omega }\times Y}\partial _{y_{i}}(y_{j}\partial _{x_{j}}u^{0})%
\text{ }v_{i}\text{ }dxdy\Big]+O(\varepsilon ).
\end{eqnarray*}%
Since $v$ vanishes on all boundaries,
\begin{eqnarray*}
\Psi &=\sum_{i=1}^{n}\Big[& \int_{\widetilde{\Omega }\times
Y}\partial _{x_{i}}u^{0}v_{i}\;dxdy+\int_{\widetilde{\Omega }\times
Y}\partial
_{y_{i}}u^{1}v_{i}\;dxdy \\
&& +\sum_{j=1}^{n}\int_{\widetilde{\Omega }\times Y}\partial
_{y_{i}}(y_{j}\partial _{x_{j}}u^{0})\,v_{i}\;dxdy \\
&& -\sum_{j=1}^{n}\int_{%
\widetilde{\Omega }\times Y}\partial _{y_{i}}(y_{j}\partial
_{x_{j}}u^{0})\,v_{i}\text{ }dxdy\Big]+O(\varepsilon ),
\end{eqnarray*}%
or after simplification,
\begin{equation*}
\Psi =\sum_{i=1}^{n}\Big[\int_{\widetilde{\Omega }\times Y}\partial
_{x_{i}}u^{0}v_{i}\;dxdy+\int_{\widetilde{\Omega }\times Y}\partial
_{y_{i}}u^{1}v_{i}\;dxdy\Big]+O(\varepsilon ).
\end{equation*}%
Finally, thanks to linearity of integral and by factoring $v_{i},$%
\begin{equation*}
\Psi =\sum_{i=1}^{n}\Big[\int_{\widetilde{\Omega }\times Y}(\partial
_{x_{i}}u^{0}+\partial _{y_{i}}u^{1})v_{i}\;dxdy\Big]+O(\varepsilon
).
\end{equation*}
\end{itemize}


%% file: GradientProof/diffusion_equation.tex
We detail the proof of Proposition \ref{model:approx:prop}. We start
with test functions $v^{0}\in \mathcal{C}_{0}^{\infty
}(\widetilde{\Omega })$ and
$v^{1}\in \mathcal{C}^{\infty }(\widetilde{\Omega }\times Y)$ that is $Y$%
-periodic.

\begin{itemize}
\item \textbf{Step 1.} We choose $v=B(v^{0}+\varepsilon v^{1})$ as a test
function in the weak formulation
(\ref{model:diffusion:variational:eq}) of the model,
\begin{equation*}
\sum_{i=1}^{n}\int_{\Omega }a(\partial _{x_{i}}u)\,\partial
_{x_{i}}B(v^{0}+\varepsilon v^{1})\;dx=\int_{\Omega
}f\,B(v^{0}+\epsilon v^{1})\;dx.
\end{equation*}

\item \textbf{Step 2.} Applying the rule (\ref{B:derivative}) of partial
derivatives of $B(v)$ yields%
\begin{equation*}
\sum_{i=1}^{n}\int_{\Omega }a(\partial _{x_{i}}u)\;(B(\partial
_{x_{i}}(v^{0}+\varepsilon v^{1}))+\frac{1}{\varepsilon }B(\partial
_{y_{i}}(v^{0}+\varepsilon v^{1}))\;dx=\int_{\Omega
}f\,B(v^{0}+\epsilon v^{1})\;dx.
\end{equation*}

By linearity of $\partial $ and $B$, and since $v^{0}$ does not depend on $y$%
, we get after application of (\ref{O3}) and simplifications,%
\begin{equation*}
\sum_{i=1}^{n}\int_{\Omega }a(\partial _{x_{i}}u)\;(B(\partial
_{x_{i}}v^{0})+B(\partial _{y_{i}}v^{1}))\;dx=\int_{\Omega
}f\,B(v^{0})\;dx+O(\varepsilon ).
\end{equation*}%
From the linearity of $B$ again,%
\begin{equation*}
\sum_{i=1}^{n}\int_{\Omega }a(\partial _{x_{i}}u)\;B(\partial
_{x_{i}}v^{0}+\partial _{y_{i}}v^{1})\;dx=\int_{\Omega
}f\,B(v^{0})\;dx+O(\varepsilon ).
\end{equation*}

\item \textbf{Step 3.} The zero-order approximation (\ref{B:approximation:1}%
) of $B$ by $T^{\ast }$ implies%
\begin{equation*}
\sum_{i=1}^{n}\int_{\Omega }a(\partial _{x_{i}}u)\;T^{\ast
}(\partial _{x_{i}}v^{0}+\partial _{y_{i}}v^{1})\;dx=\int_{\Omega
}f\,T^{\ast }(v^{0})\;dx+O(\varepsilon ).
\end{equation*}

\item \textbf{Step 4.} Now, we apply the definition of the adjoint $T^{\ast
} $ of $T$,%
\begin{equation*}
\sum_{i=1}^{n}\int_{\widetilde{\Omega }\times Y}T(a\partial
_{x_{i}}u)\;(\partial _{x_{i}}v^{0}+\partial _{y_{i}}v^{1})\;dxdy=\int_{%
\widetilde{\Omega }\times Y}T(f)\,v^{0}\;dxdy+O(\varepsilon ).
\end{equation*}

\item \textbf{Step 5.} From the identity (\ref{T:property}) and the
assumptions (\ref{model:diffusion:assumption}) we get%
\begin{equation*}
\sum_{i=1}^{n}\int_{\widetilde{\Omega }\times Y}a^{0}\,T(\partial
_{x_{i}}u)\;(\partial _{x_{i}}v^{0}+\partial _{y_{i}}v^{1})\;dxdy=\int_{%
\widetilde{\Omega }\times Y}f^{0}\,v^{0}\;dxdy+O(\varepsilon ).
\end{equation*}

\item \textbf{Step 6.} From the approximation (\ref{gradient:approxim:eq})
of the derivative operator applied to the test function
$a^{0}(\partial _{x_{i}}v^{0}+\partial _{y_{i}}v^{1})$ we get the
wanted two-scale model
\begin{equation*}
\sum_{i=1}^{n}\int_{\widetilde{\Omega }\times Y}a^{0}\,(\partial
_{x_{i}}u^{0}+\partial _{y_{i}}u^{1})\;(\partial
_{x_{i}}v^{0}+\partial _{y_{i}}v^{1})\;dxdy=\int_{\widetilde{\Omega
}\times Y}f^{0}\,v^{0}\;dxdy+O(\varepsilon ).
\end{equation*}
\end{itemize}
